\documentclass[10pt,english,journal]{IEEEtran}
\usepackage[T1]{fontenc}
\usepackage[latin9]{inputenc}
\usepackage{geometry}
\usepackage{amsmath}
\usepackage{xpatch}
\usepackage{amssymb}
\usepackage{esint}
\usepackage{mathtools}
\usepackage{amsthm}
\usepackage{nicefrac}
\usepackage{bigints}
\usepackage{mathtools}
\usepackage{blkarray, bigstrut}
\usepackage{physics}
\usepackage{calligra}
\usepackage{graphicx}
\usepackage{epstopdf}
\usepackage{dsfont}
\usepackage{array,ragged2e}
\usepackage{enumitem}
\usepackage{etoolbox}
\usepackage{babel}
\usepackage[nopar]{lipsum}
\usepackage{psfrag}
\usepackage[ruled,lined,linesnumbered]{algorithm2e}
\usepackage{algorithmic}
\usepackage{algorithm2e}
\usepackage{balance}
\usepackage{float}
\usepackage{tabularx}
\usepackage{hyperref}
\usepackage{xcolor}
\usepackage{pbox}
\usepackage{rotating}

\usepackage{tikz}

\usepackage{subcaption}
\theoremstyle{definition}

\SetKw{KwBy}{by}

\makeatletter
\newcommand{\removelatexerror}{\let\@latex@error\@gobble}
\makeatother

\graphicspath{ {Figures/} }

\newcommand{\rev}[1]{{\color{black}{#1}}}

\definecolor{lightgraybg}{RGB}{240, 240, 240}

\theoremstyle{definition}

\xpatchcmd{\proof}{\hskip\labelsep}{\hskip5\labelsep}{}{}  
\makeatletter
\xpatchcmd{\proof}{\@addpunct{.}}{\@addpunct{:}}{}{}
\makeatother

\renewcommand\[{\begin{equation}}
\renewcommand\]{\end{equation}} 
\usepackage{listings}
\usepackage{fancyvrb}
\usepackage{framed}

\usepackage{courier}
\usepackage[usenames,dvipsnames,table]{xcolor}

\definecolor{dkgreen}{rgb}{0,0.3,0}
\definecolor{gray}{rgb}{0.5,0.5,0.5}

\definecolor{codegreen}{rgb}{0,0.6,0}
\definecolor{codegray}{rgb}{0.5,0.5,0.5}
\definecolor{codepurple}{rgb}{0.58,0,0.82}
\definecolor{backcolour}{rgb}{0.95,0.95,0.92}
\definecolor{skyblue}{RGB}{135,206,235}
\lstdefinestyle{mystyle}{
    backgroundcolor=\color{backcolour},  
    commentstyle=\color{codegreen},      
    keywordstyle=\color{magenta},        
    numberstyle=\tiny\color{codegray},   
    stringstyle=\color{codepurple},      
    basicstyle=\ttfamily\footnotesize,   
    breakatwhitespace=false,             
    breaklines=true,                     
    captionpos=b,                        
    keepspaces=true,                     
    numbers=left,                       
    numbersep=5pt,                      
    showspaces=false,                   
    showstringspaces=false,             
    showtabs=false,                     
    tabsize=2                           
}

\makeatletter
\newcommand*{\rom}[1]{\expandafter\@slowromancap\romannumeral #1@}
\makeatother

\lstdefinestyle{promptstyle}{
    backgroundcolor=\color{gray!10}, 
    commentstyle=\color{green!50!black}, 
    keywordstyle=\color{blue}, 
    numberstyle=\tiny\color{gray}, 
    stringstyle=\color{purple}, 
    basicstyle=\ttfamily\small, 
    breakatwhitespace=false, 
    breaklines=true, 
    captionpos=b, 
    keepspaces=true, 
    numbers=left, 
    numbersep=5pt, 
    showspaces=false, 
    showstringspaces=false, 
    showtabs=false, 
    tabsize=2, 
    rulesepcolor=\color{gray!20} 
}

\usepackage{siunitx}
\usepackage{tabu}
\usepackage{booktabs}
\usepackage{multirow}
\usepackage{capt-of}
\usepackage{array}
\usepackage{arydshln}
\usepackage{cite}
\usepackage{url} 

\newcommand{\comment}[1]{}

\newcommand{\revi}[1]{{\color{black}{#1}}}

\makeatletter
\patchcmd{\@maketitle}
  {\addvspace{0.5\baselineskip}\egroup}
  {\addvspace{-1\baselineskip}\egroup}
  {}
  {}
\makeatother

\begin{document}

\title{A Tutorial on Cognitive Biases in Agentic AI-Driven 6G Autonomous Networks}

\author{
Hatim~Chergui,~\IEEEmembership{Senior~Member,~IEEE,} Farhad~Rezazadeh,~\IEEEmembership{Member,~IEEE,} Merouane~Debbah,~\IEEEmembership{Fellow,~IEEE} \\and Christos~Verikoukis,~\IEEEmembership{Senior~Member,~IEEE}

\IEEEcompsocitemizethanks{\IEEEcompsocthanksitem H. Chergui is with the i2CAT Foundation, Spain. (e-mail: chergui@ieee.org)}
\IEEEcompsocitemizethanks{\IEEEcompsocthanksitem F. Rezazadeh is with the Hostelworld Group, BrainOmega, and the Technical University of Catalonia (UPC), 08860 Castelldefels, Spain (e-mail: farhad.rezazadeh@upc.edu).}
\IEEEcompsocitemizethanks{\IEEEcompsocthanksitem M. Debbah is with the Khalifa University of Science and Technology, 127788 Abu Dhabi, UAE (e-mail: merouane.debbah@ku.ac.ae).}
\IEEEcompsocitemizethanks{\IEEEcompsocthanksitem C. Verikoukis is with ISI/ATH and University of Patras, Greece. (e-mail: cveri@isi.gr)}}
\maketitle

\begin{abstract}
\revi{The path to higher network autonomy in 6G lies beyond the mere optimization of key performance indicators (KPIs), requiring systems that perceive and reason over the network environment \emph{as it is}. This can be achieved through \emph{agentic AI}, where large language model (LLM)-powered agents utilize multimodal telemetry, memory, and cross-domain negotiation to achieve multi-objective goals. However, deploying such agents introduces cognitive biases inherited from human design, which can severely distort reasoning and actuation. This paper provides a comprehensive tutorial on well-known cognitive biases, detailing their taxonomy, mathematical formulation, emergence in telecom systems, and tailored mitigation strategies. We validate these concepts through two distinct use-cases in 6G management\footnote{\revi{The source codes of these use cases are available for non-commercial use at: https://github.com/HatimChergui/}}. First, we tackle \emph{anchoring bias} in inter-slice resource negotiation. To overcome the prohibitive execution delays of cloud-based LLMs, this use-case deploys a locally hosted 1B-parameter model on an RTX A4000 GPU, successfully achieving sub-second inference latencies compatible with near-real-time (near-RT) operations. By replacing fixed heuristic anchors with a Truncated Weibull randomized anchor strategy, the agents dismantle rigid biases, intelligently consume SLA slack, and dynamically double the system-wide energy savings (peaking at 25\%) without violating strict latency limits. Second, we mitigate \emph{temporal and confirmation biases} in RAN-Edge cross-domain negotiation by designing an unbiased collective memory. By integrating semantic/temporal decay and an inflection bonus that actively highlights past negotiation failures, agents are prevented from over-relying on recent data or repeating past mistakes. Grounding decisions in this richer, debiased historical context yields highly robust agreements, achieving a $\times 5$ latency reduction and roughly $40\%$ higher energy savings compared to memoryless baselines.}
\end{abstract}

\maketitle

\section{Introduction}

\IEEEPARstart{T}{he} landscape of telecom automation over the past 15 years can be structured along TM Forum's maturity levels, from Level 1 (script-based) to Level 3 (analytics- and AI-assisted orchestration) \cite{tmforum2021autonomous, Newref}, with initiatives such as 3GPP Self-Organizing Networks (SON), ETSI Zero-Touch Service Management (ZSM) and O-RAN. While their introduced paradigms have yielded operational efficiency, their optimization logic remains narrowly tethered to KPIs. However, those metrics are imperfect proxies/surrogates for the underlying objectives of communication systems. This reliance induces a well-documented failure mode captured by \textit{Goodhart's Law}, which states that ``when a measure becomes a target, it ceases to be a good measure'' \cite{goodhart}. In practice, this manifests as \textit{metric fixation} \cite{muller}, where algorithmic optimization maximizes numerical scores while neglecting harder-to-measure essence of the telecom environment and its users. In anticipation of 6G, overcoming this proxy-goal misalignment necessitates a transition from metric-centric optimization to \textit{goal-directed autonomy}. This implies not only moving towards TM Forum Levels 4 and 5, but also embedding intelligence that can operate across dynamic, multi-domain, and partially observable environments. \textit{Agentic AI} provides a viable paradigm: LLM-powered autonomous agents endowed with the ability to directly perceive through multimodal data (without proxies), reason over historical and contextual knowledge, plan adaptive courses of action, communicate under diverse intents, and act through APIs or network control functions. Crucially, such agents can negotiate across domains (e.g., RAN, edge, and core) to reconcile competing objectives, adapt to unforeseen conditions, and iteratively refine strategies via explanation and feedback. In this way, agentic AI would serve as the cognitive substrate for 6G networks, bridging the gap between low-level signals and high-level semantic goals. Rather than optimizing proxies, networks can pursue objectives that more faithfully represent the essence of telecommunication.
\subsection{Cognitive Biases in Agentic Systems}
Special attention should be paid, however, to the susceptibility of agentic systems to the so-called \emph{cognitive biases}, which are \emph{systematic deviations from rational judgment;} a concept originally rooted in human psychology that agents could inherit in various ways and perpetuate, posing a tangible threat to the efficiency, fairness, and reliability of next-generation autonomous networks. Biases affect every layer of an agent's behavior, from initial perception and reasoning to final decision-making and action execution. The foundational work on cognitive biases in human judgment was established by Tversky and Kahneman in their seminal paper, ''Judgment under Uncertainty: Heuristics and Biases'' \cite{Tversky1974}. Their research demonstrated that humans rely on cognitive shortcuts, or heuristics, which can lead to predictable and systematic errors in judgment. In agentic systems, these biases are inherited and manifested across multiple layers. The insidious impact of biases can be observed throughout the entire agentic system pipeline, from perception to action. i) \emph{Prompts} are a primary entry point for bias, where framing effects or the specific wording of instructions can skew an agent's perception of a problem. For instance, prompting a network agent to ``maximize throughput at all costs'' can lead to a solution that ignores critical latency or fairness metrics. ii) Biases are deeply embedded in the \emph{training data,} stemming from historical imbalances, cultural skews, and sampling errors. A large language model trained on data from a region with a specific network infrastructure could, when deployed elsewhere, exhibit a bias towards that legacy architecture, failing to optimally manage a new, more advanced one. Furthermore, biases emerge and compound in an agent's internal processes, with implications for its reasoning and tool use. iii) The agent's \emph{reasoning} paths are susceptible to various heuristics. An agent tasked with dynamic resource allocation could exhibit an availability heuristic, disproportionately allocating bandwidth to a specific network slice from which it has received the most recent or frequent requests, thereby neglecting other slices in need. Similarly, a security agent may fall prey to a confirmation bias, only seeking out evidence that confirms a pre-existing threat model and overlooking a novel, unseen attack vector. These behaviors are akin to human shortcuts, where evidence is mis-weighted and readily accessible information is prioritized. iv) Biases are also evident in \emph{tool use,} which includes the way memories are stored and retrieved, and how data sources and APIs are selected. An agent's memory retrieval might be subject to recency/primacy biases, causing it to favor recently processed network logs over a more complete historical record, leading to short-sighted decisions. Conversely, an authority bias could lead an agent to show a strong, uncritical preference for data from a single, \emph{authoritative} source or a familiar API, even when more suitable or diverse tools are available. The recent work by Xie et al. \cite{Xie2024} highlights the growing concern of these biases, exploring their manifestation in large language models within multi-agent systems, a structure increasingly relevant to decentralized 6G architectures.

\rev{
Note that agentic biases generally originate from common mechanisms across learning-based systems, including skewed training data, imperfect objectives, model inductive biases, and partial or noisy observations. In this sense, the emergence of biases in various other areas is not fundamentally different from that in LLM agents envisioned for 6G networks. The main distinction lies in how these biases manifest: in other areas, such as healthcare or finances, they typically affect semantic reasoning and language outputs, whereas in 6G agents they appear as systematic errors in control or resource allocation policies derived from biased measurements or traffic data. Therefore, the underlying mechanisms are largely shared across domains, while the domain-specific state, action, and evaluation spaces determine how biases are expressed and detected.

}

\subsection{Related Work}
Drawing parallels with well-documented human biases such as confirmation bias, recency effects, and groupthink, recent research examines how similar distortions can manifest both within individual AI agents and across their interactions, potentially affecting collective decision-making, fairness, and safety. A comprehensive contribution in this domain is \emph{MindScope} \cite{mindscope}, which introduces a dataset encompassing 72 human cognitive biases and leverages it to evaluate how LLM agents display these distortions during multi-agent dialogue. Findings indicate that even when models appear well-aligned in isolation, their interactions can trigger latent, higher-order biases such as anchoring effects. Mitigation strategies---including retrieval-augmented generation (RAG), structured debate mechanisms, and reinforcement learning (RL)-based adjudication---have shown promise in reducing these risks.  

Expanding on this, Liu et al.~\cite{unmasking} investigate the emergence of conversational echo chambers in multi-agent settings. Their work demonstrates that iterative discussions often amplify existing biases, as agents gradually converge toward consensus-seeking positions, which is an emergent phenomenon not observed in single-agent contexts. This underscores that distortions can originate from interaction dynamics themselves, rather than solely from pretrained knowledge.  

In a more structural way, the \emph{fairness in agentic AI} framework \cite{fairness} examines how systemic distortions may arise through decentralized collaboration among agents. The framework connects ethical alignment with incentive structures and negotiation constraints, offering design tools to mitigate emergent bias and to prevent inequitable treatment of users or tasks embedded in implicit reward mechanisms. Complementarily, the \emph{hidden profile benchmark} \cite{assessing} adapts a classic social psychology paradigm to multi-agent LLMs, revealing that agents often fail to uncover critical but unevenly distributed information held by peers-mirroring groupthink and informational bias in human teams. This highlights fundamental vulnerabilities in reasoning diversity and emphasizes the need for diversity-aware memory structures and communication protocols.  

Having said that, a tutorial introduction to cognitive biases in the context of agentic AI-based 6G network automation is yet to be considered, examining their emergence, affected agentic components, practical 6G examples and mitigation strategies.

\subsection{Contributions}
This paper makes the following contributions:  
\begin{itemize}
    \item Presents an overview of agentic systems for telecom, with their components, protocols and interactions as well as reasoning and planning capabilities while also briefly reviewing literature in this emerging area.  
    \item Develops a systematic tutorial on cognitive biases and their emergence, including taxonomy, impact on the various agentic components such as reasoning, planning, memory, negotiation, tool use, and actuation, as well as provides practical 6G examples and mitigation approaches.  
    \item Demonstrates the emergence and implications of these biases in 6G management scenarios, with a focus on RAN-Edge cross-domain orchestration and service level agreement (SLA)-sensitive optimization.  
    \item Illustrates the mitigation gain of some well-known biases through two use-cases on 6G inter-slice and cross-domain agentic negotiation. Specifically, debiasing mechanisms targeting negotiation as well as memory use are explored to encourage bolder decisions and avoid anchoring, temporal and confirmation biases, wherefore the resulting gains are assessed in terms of latency and energy saving. 
      
\end{itemize}

\section{Agentic AI-Driven Network Automation}

\subsection{Typical Components of an Agentic System}
\label{sec:agentic-components}

A modern telecom agentic system is a tightly integrated, multi-layered architecture whose components collectively enable perception, semantic reasoning, secure negotiation, safe actuation and continual learning as depicted in Figure \ref{fig:comp}. The following subsections describe the canonical elements, their responsibilities, and the integration patterns that turn a collection of tools into a coherent agentic fabric suitable for various technological domains.

\subsubsection{\textbf{LLM-empowered Agent}}
The core agent is an LLM-centered \cite{Friha2024LLMBasedEI,Picano2025AMG} reasoning and control entity that orchestrates the perception-reason-plan-act loop. Beyond natural-language fluency, it implements modular capabilities: a perception front-end that ingests latent representations, a retrieval-augmented memory interface, a symbolic/planning module that generates candidate strategies, a negotiator that composes intentful messages for peers, a verifier that queries a digital twin for validation, and an executor that issues guarded network API calls. Agents must support stateful dialogues, multi-turn negotiation, and constrained action synthesis (e.g., produce candidate plans with cost/risk annotations). Practically, the agent exposes two main kinds of interfaces: (i) tool invocation via an MCP-like function call semantics (e.g., \texttt{mcp.call("digital\_twin.simulate", scenario={...}, horizon=30s)}), and (ii) agent-to-agent messages using an A2A protocol carrying structured intents (proposal, counter-proposal, justification, commit).

\subsubsection{\textbf{Digital Twin and Prediction Tools}}
A high-fidelity digital twin (DT) functions as the agent's safe playground \cite{jiang2024links}: agents submit what-if scenarios, stress tests, or short rollouts to evaluate the temporal and cross-domain impact of candidate actions before committing them to live infrastructure. Twins range from lightweight surrogate models that provide millisecond feedback to detailed simulators for deep validation. Key responsibilities include maintaining synchronized state with production (shadowing), supporting rollback/canary experiments, exposing cost models and failure modes, and providing counterfactual traces that agents use to attach confidence and expected utility to proposals. Prediction Services \cite{Rezazadeh2025RivalingTM} also allow agents to perform farsighted decisions.
\begin{figure}
    \centering
    \includegraphics[width=1.0\linewidth]{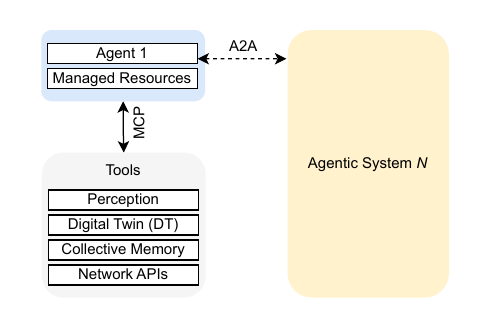}
    \caption{Typical components of a 6G agentic system.}
    \label{fig:comp}
\end{figure}
\subsubsection{\textbf{Memory and Experience Store}}
Memory enables sample-efficient reasoning and faster consensus by allowing agents to recall past strategies, observed outcomes, and negotiated compromises \cite{unbiased}. Architecturally, memory comprises multiple layers: short-term (session-level) buffers for ongoing negotiations, episodic records of full negotiation-execution episodes, and a semantic/latent index (vector store) for rapid similarity search. Basic retrieval policies prioritize recent, high-reward episodes and semantically matching contexts. Memory interfaces must support semantic queries (e.g., embedding and approximate nearest neighbor (ANN) lookup), provenance metadata, and secure access controls so agents can justify decisions with human-interpretable precedents.

\subsubsection{\textbf{Data Sources and Perception Stack}}
Robust perception aggregates heterogeneous telemetry, e.g., time-series KPIs, protocol traces, topology and inventory, user-plane metrics, and multi-modal signals (e.g., maps, logs). A preprocessing pipeline performs schema harmonization, normalization, feature extraction and embedding into latent spaces suited for LLM agents and downstream planners. Event detection, anomaly scoring, and context windows feed the agent's situational awareness. Critically, the perception stack must provide uncertainty estimates and data-quality indicators so reasoning modules can weigh observations correctly.

\subsubsection{\textbf{Network APIs and Actuation}}

\rev{Actuation in agentic AI for telecom networks is performed through domain-specific APIs that expose fine-grained control over the RAN, edge services, core functions, and orchestration platforms. For instance, an srsRAN API may allow adjusting scheduling parameters or PRB allocations, while a Kubernetes (K8s) API can scale edge services or redeploy VNFs. 

To better understand the capacity and reliability of underlying platforms such as Kubernetes, OpenStack, and Docker, agents must consider several key technical dimensions, including compute and memory resource availability, network throughput and latency, container or VM lifecycle management, orchestration and scheduling behavior, and fault tolerance mechanisms such as replication and automated recovery. These dimensions determine how reliably services can scale, migrate, or recover from failures during automated actuation.

Safety mechanisms are therefore critical. A proposed configuration change (e.g., a new QoS profile on a 5QI bearer or scaling of an edge service in K8s) should first be validated in a digital twin or staged on a subset of pods before system-wide rollout. Agents should also avoid performing irreversible operations without a validation step, where human approval may be required for actions such as core network function restarts. Finally, audit logs, execution receipts, and telemetry feedback must be systematically captured so that the agent's memory can be updated with ground-truth outcomes (e.g., whether a predicted CPU scaling action in K8s actually reduced latency), closing the loop between reasoning, action, and verified system state.}

\subsubsection{\textbf{Communication Protocols, Ontologies and Agent Fabric}}
Inter-agent coordination rests on well-defined communication protocols \cite{rosenberg_jennings_2025_ai_agent_protocols} and shared ontologies. An agent-to-agent (A2A) protocol carries structured messages with typed fields (e.g., intent, proposals, reason). Several intents exists, e.g., \emph{propose, counter-propose, query, explain, confirm, reject, commit.} In practice, the so-called \emph{agent fabric} provides the runtime substrate where multiple agents are deployed, coordinated, and scaled across distributed environments. It offers common services such as discovery, messaging, state sharing, and lifecycle management, enabling agents to communicate and operate collectively rather than in isolation. The Model Context Protocol (MCP) or equivalent tool-invocation channels, on the other hand, permit standardized tool calls and return structured outputs. Semantic encodings (ontologies, schema registries, and vocabularies) are essential so proposals and justifications are machine-interpretable across vendors and domains. The TM Forum Intent Ontology (TIO), for instance, provides a standardized semantic layer for expressing intents in network and service management, ensuring that agent messages carry domain-aware meaning. By aligning inter-agent proposals and actions with TIO, heterogeneous agents can interpret, decompose, and fulfill intents consistently across RAN, edge, and core domains. This ontology-driven alignment enhances interoperability and automation by grounding agent reasoning in a common industry-backed vocabulary.

\subsection{Agentic Reasoning and Planning in 6G}
\label{subsec:agentic-reasoning-planning}

We focus on LLM-powered agents that reason over multimodal telecom context and synthesize safe, verifiable action plans spanning RAN, edge, and core. Reasoning treats the network as a belief-driven stochastic environment, while planning composes constrained, tool-grounded action sequences that must be validated in a digital twin (DT) before actuation. To stay consistent with Fig.~\ref{fig:comp} and the bias taxonomy in Table~\ref{tab:bias-taxonomy}, we first formalize the reasoning objective, then connect it to how models are trained to reason, and finally describe specialization to 6G data and interfaces.

\textbf{Definition:}(Reasoning in Agentic 6G)
Given observations $x$ (telemetry, logs, intents), background knowledge $\mathcal{K}$ (standards, policies, ontologies), and tools $\mathcal{T}$ (DT, telemetry services, network APIs), an agent seeks a plan $\pi=(a_1,\ldots,a_H)$ from a \emph{safe action set} $\mathcal{A}_{\mathrm{safe}}$ that maximizes a multi-objective utility $U(\pi) \in \mathbb{R}^m$ (e.g., latency, energy, fairness, risk), under operational and chance constraints:
\[
\max_{\pi \subset \mathcal{A}_{\mathrm{safe}}} \ \mathbb{E}[U(\pi)\mid x,\mathcal{K}]
\quad \text{s.t.}\quad
\Pr\{g_j(\pi,x)\le 0\}\ge 1-\epsilon_j,\ \forall j.
\]

\subsubsection{\textbf{Logical Reasoning Types in 6G Agents}}
Logical reasoning manifests in several complementary forms. \emph{Deductive reasoning} maps high-level intents and policies to concrete, correctness-preserving configurations (e.g., from an intent to a 5QI profile and scheduler settings), providing compliance and safety guarantees. \emph{Inductive reasoning} \cite{Chen2025ASO} generalizes from historical episodes or DT rollouts to propose candidate actions with calibrated confidence, for example, inferring traffic--latency relationships that inform proactive scaling. \emph{Abductive or causal reasoning} explains observed degradations by positing plausible causes and minimal interventions (e.g., distinguishing backhaul congestion from CPU throttling as the source of latency spikes) and selecting actions with the highest expected counterfactual gain. \emph{Analogical reasoning} recognizes structural similarity between a novel context and past episodes---such as slice mix, RF conditions, or policy constraints---to adapt or compose previously successful strategies under current constraints. Finally, \emph{counterfactual/model-based reasoning} \cite{Khadilkar2025CausalCounterfactualRT} uses the DT to explore ``what--if'' scenarios and compare interventions before committing changes to live infrastructure. In practice, agents combine these modes within a single loop: hypotheses are generated abductively, stress-tested counterfactually, and distilled into deductively checkable configurations that can be generalized inductively or by analogy across sites and time.

\subsubsection{\textbf{From Pretraining to Operational Reasoning}}
Modern agents learn these behaviors in stages. During broad pretraining on text, code, and quantitative material (e.g., GSM8K maths dataset), LLMs acquire competence in symbolic manipulation and intermediate-step fidelity, which supports the derivation and verification of multi-step rationales \cite{gsm8k}. Instruction-style fine-tuning then aligns the model to follow domain procedures, call tools when needed, and expose intermediate steps transparently, rather than relying on brittle internal shortcuts. In operation, self-consistency sampling and external verification further stabilize reasoning. The agent generates diverse rationales, selects coherent ones, and validates critical calculations and predictions using the DT or specialized analyzers. This separation between ``thinking'' and ``checking'' reduces single-path failures and yields uncertainty estimates that planning can consume.

\subsubsection{\textbf{Telecom Specialization and Domain Grounding}}
Specializing for 6G requires grounding the agent in domain-structured data and APIs. Inputs and outputs are normalized using a canonical KPI schema and shared ontologies, ensuring quantities, units, and timestamps are comparable across domains; this reduces framing effects and prevents spurious differences caused by formatting. Supervised episodes pair observations with concise rationales and concrete actions (telemetry $\rightarrow$ explanation $\rightarrow$ API calls), curated from traces and DT rollouts. Crucially, both successes and failures are included to avoid survivorship bias and to teach the boundaries where strategies break down. Preference-style optimization ranks plans by multi-objective utility and safety violations, encouraging Pareto-efficient behavior and penalizing hallucinated tool use. Finally, uncertainty-aware generation produces plans with explicit risk bounds, which are stress-tested under traffic and RF perturbations in the DT before any limited-scope canary actuation.

\subsubsection{\textbf{Constrained Plan--Verify--Act Loop}}
The operational loop follows \emph{reason--plan--verify--act}. Reasoning fuses live telemetry, policies, and retrieved episodes to hypothesize causes and propose interventions, emitting intermediate steps with calibrated uncertainty. Planning then solves a constrained rollout over $\mathcal{A}_{\mathrm{safe}}$; when constraints dominate (e.g., strict URLLC guarantees), lexicographic priorities ensure safety before efficiency. Verification executes counterfactuals in the DT to compute cost vectors and feasibility; plans with low confidence or high regret are rejected or revised. Actuation performs guarded API calls with staging or canaries, and outcomes are written back to episodic memory for continual improvement.

\subsubsection{\textbf{Bias-aware Hooks During Reasoning and Planning}}
Because reasoning and planning are susceptible to systematic distortions, we insert lightweight, auditable hooks that align with Table~\ref{tab:bias-taxonomy}. Counterfactual checks in the DT encourage consideration of disconfirming alternatives, thereby mitigating confirmation, availability, and anchoring effects. Diversity-aware retrieval balances recent and older evidence, successes and failures, and semantic variety to curb recency/primacy and survivorship. Canonical KPI representations and neutralized prompts reduce framing and suggestion biases by standardizing inputs before inference. Uncertainty calibration propagates confidence throughout the pipeline to counter neglect of probability and automation bias. Finally, source-oblivious tool comparison hides vendor or authority labels during selection and demands dual-source consistency before commit, mitigating authority and halo effects.

\subsubsection{\textbf{Worked Formulation}}
Let $a\in\mathcal{A}_{\mathrm{safe}}$ denote a joint RAN--edge action (e.g., PRB split and CPU frequency). The DT returns a cost vector $\mathbf{c}(a) = [L(a), E(a), F(a), R(a)]$ for latency, energy, fairness, and risk. A typical objective is
\begin{equation}
\begin{aligned}
\min_{a\in\mathcal{A}_{\mathrm{safe}}} \ & \mathbf{w}^\top \mathbf{c}(a) \\
\text{s.t.} \quad 
& \Pr\{L(a)\le L_{\mathrm{SLA}}\}\ge 1-\epsilon, \\
& R(a)\le R_{\max},
\end{aligned}
\label{eq:optimization}
\end{equation}
with candidates ranked by expected regret against a baseline and by feasibility across sampled traffic/RF scenarios. Non-dominated plans are surfaced for negotiation or executed via canary rollout, with post-actuation feedback to close the loop.

\subsection{Agentic Systems in 6G: A Review}

Recent studies have explored the use of agentic AI for autonomous network management, each advancing the field from different perspectives. In \cite{unbiased}, the authors consider a RAN-Edge cross domain management scenario where they introduce a new architecture of the long-term agentic memory to efficiently retrieve relevant stored collective agreements of the RAN and Edge LLM-powered agents and capitalize on them, resulting in lower negotiation failures and SLA violation as well as improved latency and energy saving. Ferrag et al. \cite{ferrag2025llmreasoningautonomousai} present a comprehensive survey of agentic AI, covering approximately 60 benchmarks across reasoning, planning, tool use, multimodal and embodied tasks, interactive workflows, and agent orchestration, while also proposing a taxonomy of LLM-agent frameworks. Their work further reviews emerging agent-to-agent communication protocols such as the Agent Communication Protocol, Model Context Protocol, and A2A, and highlights key limitations, including failure modes, security risks, and open challenges. In parallel, Chen et al. \cite{chen} introduce BlockAgents, which integrate blockchain consensus and proof-of-thought mechanisms into LLM-based multi-agent coordination to ensure Byzantine-robust role assignment, proposal evaluation, and decision-making. Taken together, these studies reflect a broader convergence in the literature around three main directions: the development of modular agent architectures that integrate reasoning, memory, and tool use; the design of collaborative protocols for negotiation, consensus, and hierarchical control; and the incorporation of robustness and trust mechanisms, such as decentralized consensus and auditing, to safeguard against adversarial or faulty agents. On the other hand, \cite{zou2023wirelessmultiagentgenerativeai} introduce the concept of wireless multi-agent generative AI, positioning LLM-based agents as enablers of a transition from connected to collective intelligence in wireless networks. They analyze decentralized cooperation and competition through a game-theoretic lens, propose architectural principles for deploying on-device agents, and illustrate their vision with intent-based networking as a case study. Besides, \cite{xu2024largelanguagemodelagents} focuses on integrating LLM agents into 6G networks via a modular framework of perception, grounding, and alignment, distributed across devices and edge servers. They propose a split learning paradigm and model caching strategies to optimize resource allocation and coordination, enabling applications such as integrated sensing, digital twins, and task-oriented communications.
The LINKs framework introduced in \cite{jiang2024links} constitutes one of the first attempts to integrate LLMs with digital twin-based 6G management systems, enabling agents to translate high-level intent into control actions. However, its reasoning capacity is still limited, and integration with real-time RAN telemetry is underexplored. Complementarily, \cite{xia2025agentic} presents a foundational framework that conceptualizes generative AI models as distributed agents for collaborative task planning across network elements, emphasizing emergent communication and semantic coordination, though without concrete ties to telecom protocols or operational data. A more modular approach is advanced in \cite{liu2025lameta}, which separates cloud-based intent parsing from edge-level policy generation to address scalability and latency, though validation is restricted to high-level simulations rather than real network traces. Extending beyond telecom, \cite{xu2025urban} investigates agentic digital twins for cyber-physical optimization using large generative models and MCP, demonstrating orchestration capabilities primarily in urban logistics scenarios. To enhance grounding and mitigate hallucinations, \cite{zhang2025toward} proposes a retrieval-augmented architecture that incorporates structured network knowledge into the agent reasoning loop, though this remains at the conceptual stage. Finally, \cite{tang2025edge} outlines an end-to-end design for embedding LLM agents into the O-RAN framework to automate lifecycle management of xApps and rApps, aligning well with modular RIC interfaces but still lacking empirical validation and deployment studies.

\subsection{\rev{O-RAN Components, Interfaces, and Feasibility of Bias-Mitigation Integration}}
\label{subsec:oran-summary}
\rev{
Figure~\ref{fig:oran-arch} summarizes the Open Radio Access Network (O-RAN) control architecture that we reference throughout when discussing the Radio Intelligent Controller (RIC), xApps, and the A1/E2/O1 interfaces. At the Radio Access Network (RAN) layer, O-RAN disaggregates the base station into the O-RAN Radio Unit (O-RU), O-RAN Distributed Unit (O-DU), and O-RAN Central Unit (O-CU), which align with the Third Generation Partnership Project (3GPP) New Radio (NR) / Next Generation RAN (NG-RAN) functional split; radio-frequency and low-PHY processing at the O-RU, time-critical high-PHY/MAC/RLC functions at the O-DU, and higher-layer functions at the O-CU \cite{3gpp_38300,3gpp_38401}. Above the RAN, O-RAN introduces two complementary control nodes. The Near-Real-Time Radio Intelligent Controller (Near-RT RIC) hosts near-real-time control applications (xApps) for closed-loop optimization (e.g., handover, interference, load balancing, anomaly detection) and interacts with the RAN via the E2 interface (E2), enabling telemetry subscription and bounded control actions under tight latency constraints \cite{oran_e2,oran_arch}. The Non-Real-Time Radio Intelligent Controller (Non-RT RIC) resides within the Service Management and Orchestration (SMO) layer, alongside Operations Support Systems / Business Support Systems (OSS/BSS), and operates at longer timescales for policy, analytics, model governance, and lifecycle management; it guides the Near-RT RIC through the A1 interface (A1) \cite{oran_a1,oran_arch}. Orthogonally, the O1 interface (O1) supports operations, administration, and maintenance functions (telemetry, configuration, alarms, audit), providing the long-horizon evidence channel needed for validation and accountability \cite{oran_arch}.
\begin{figure}[t]
    \centering
    \includegraphics[width=1.0\linewidth]{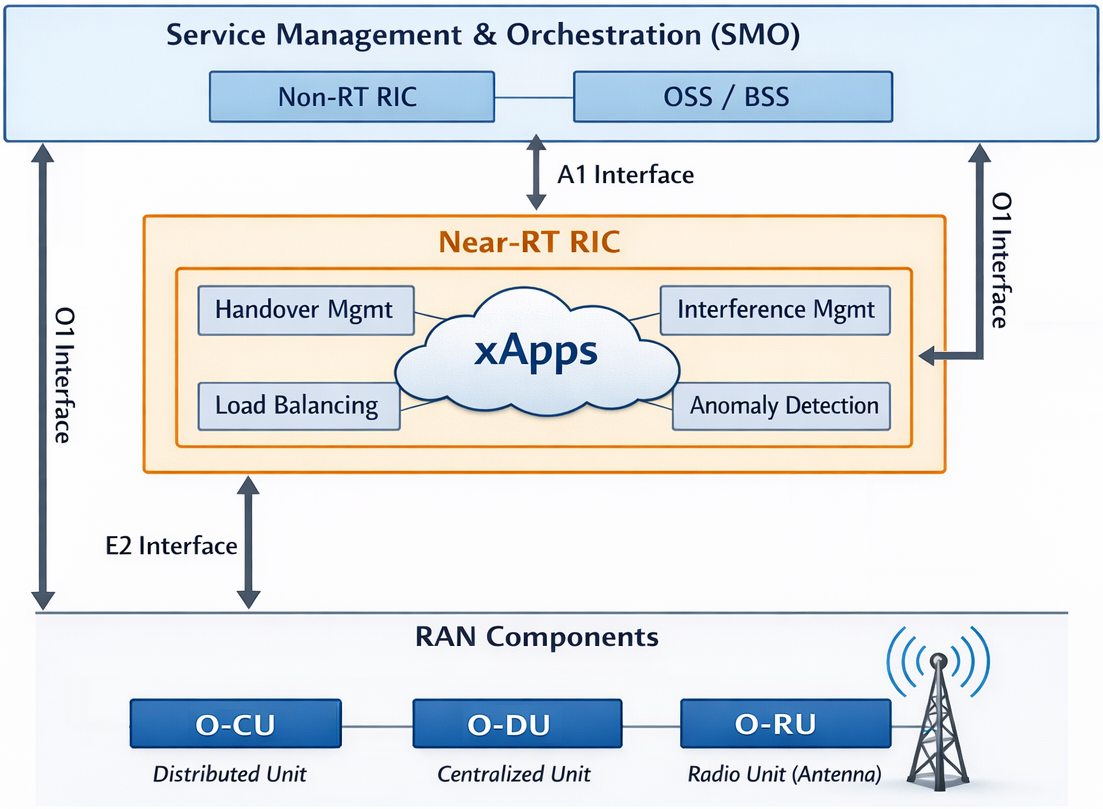}
    \caption{\rev{Reminder of a typical O-RAN architecture.}}
    \label{fig:oran-arch}
\end{figure}
This architectural separation also provides a concrete path to integrate the bias-mitigation strategies discussed in this paper into O-RAN without requiring non-standard interfaces. First, mitigation must be \emph{timescale-aligned}; methods that require broad context or multi-episode reasoning---e.g., adversarial evaluation loops, counterfactual stress testing with a digital twin, dataset balancing across successes/failures to reduce survivorship effects, or prompt/representation canonicalization to reduce framing and suggestion bias---fit naturally in SMO/Non-RT RIC because they can operate over minutes-to-hours horizons and can be expressed as policies or model governance delivered downstream via A1 \cite{oran_a1,oran_arch}. Second, mitigation must be \emph{interface-compliant}; evidence needed to challenge hypotheses (confirmation bias) or calibrate uncertainty (neglect of probability) can be obtained via E2 telemetry and complemented by O1/OAM traces for longer-window ground truth; conversely, near-real-time mitigation hooks (e.g., mandatory verification gates for high-blast-radius actions to curb automation bias, confidence thresholds and risk buffers to curb uncertainty neglect, or cross-source consistency checks to curb authority/halo effects) can be implemented inside xApps while still using the standard E2 control/indication flows \cite{oran_e2,oran_arch}. Third, mitigation must be \emph{safely deployable}; O-RAN already encourages staged control through policy constraints and OAM oversight; thus a practical plan--verify--act workflow can be realized by having Non-RT RIC/SMO enforce governance (what objectives, constraints, prompts, and retrieval rules are admissible), while Near-RT xApps execute bounded control actions only after lightweight verification using recent E2 observations and predefined safety envelopes, with all decisions logged via O1 for auditability and for closing the learning loop in the agentic memory. Under this view, the mitigation mechanisms proposed in our use cases map cleanly onto the figure; anchor randomization and negotiation protocols can be realized as xApp-level decision logic constrained by A1 policies; temporal/confirmation debiasing in collective memory is naturally hosted in SMO/Non-RT RIC as part of analytics and experience stores fed by O1/E2 telemetry; and final actuation remains standardized through E2 into the RAN components (O-CU/O-DU/O-RU). As a result, the key question is not whether bias mitigation can be integrated into O-RAN, but rather how to partition it across SMO/Non-RT versus Near-RT such that the mitigation is computationally feasible, interface-aligned, and operationally safe, while maintaining 3GPP-consistent functional splits and O-RAN-compliant control surfaces \cite{3gpp_38300,3gpp_38401,oran_arch,oran_a1,oran_e2}.

}

\begin{table*}[ht!]
\centering
\caption{\rev{Taxonomy of Cognitive Biases in LLM-Powered 6G Agents}}
\label{tab:bias-taxonomy}
\renewcommand{\arraystretch}{1.25}
\setlength{\tabcolsep}{6pt}
\scalebox{0.9}{ 
\begin{tabularx}{\textwidth}{p{3.5cm} p{5cm} p{4cm} X}
\toprule
\textbf{Bias Category} & \textbf{Definition} & \textbf{Impacted Agentic Components} & \textbf{Example of Mitigation} \\
\midrule

Confirmation Bias \cite{wason1960,Nickerson1998} & Agent selectively retrieves or interprets evidence supporting prior hypothesis. Example: xApp queries only low-load cells for PRB allocation, ignoring high-load ones. & Reasoning, Memory Queries, Telemetry Filtering & Enforce counterfactual queries or negotiation; cross-check telemetry from multiple sources before decision. \\

Recency / Primacy Bias \cite{Murdock1962,Kahneman2011} & Agent over-weighs recent or first observations. Example: Fault-recovery agent acts only on last 10 min of RSRP measurements or initial fault logs. & Memory Access, Planning & Weighted evidence fusion; validate across full telemetry timeline before reconfiguration. \\

Anchoring Bias \cite{Tversky1974,Kahneman2011} & Early proposals bias subsequent reasoning. Example: Energy optimization agent sticks to default power despite high SINR allowing lower settings. & Reasoning, Planning Initialization & Randomize anchor values; simulate multiple baseline scenarios before committing. \\

Availability Heuristic \cite{tversky1973,Kahneman2011} & Agent overestimates likelihood of salient/recent events. Example: Fault prediction overreacts to 3 recent XnAP resets, ignoring long-term failure rate. & Memory Retrieval, Risk Estimation & Normalize event likelihood using historical data; apply Bayesian priors. \\

Authority Bias \cite{Milgram1963,Kahneman2011} & Overweighting outputs from trusted sources. Example: RIC accepts vendor KPI reports without cross-checking local counters. & API Trust Assignment, Reasoning & Blind source evaluation; dual-source validation before applying config changes. \\

Halo Effect \cite{Thorndike1920,Kahneman2011} & Success in one task inflates trust in unrelated tasks. Example: xApp trusted in latency tuning because it succeeded in energy optimization. & Peer Trust Model, Reasoning & Decay trust scores across domains; cross-validate with independent measurements. \\

Suggestion / Prompting Bias \cite{Nisbett1980,Kahneman2011} & Agent influenced by wording in prompts or peer proposals. Example: prompt \emph{latency is the primary goal} disregards looking for optimal allocation. & Prompt Interpretation, API Feedback & Use neutral prompt templates; validate decisions via multi-agent cross-checks. \\

Groupthink / Herding \cite{Janis1972,Banerjee1992} & Agent combines its own weak evidence (PRB slit 1) with the signal that other agents prefer another option (PRB split 2). As a result, agent's posterior belief that 2 is the best option increases. & Multi-Agent Negotiation, Consensus & Prompt agents to generate at least one challenging proposal; simulate alternatives before consensus. \\

Framing Effect \cite{Tversky1981,Kahneman2011} & Equivalent inputs produce different decisions. Example: PRB usage framed as 20\% free vs 80\% used changes allocation. & Reasoning, Prompt Interpretation, Memory & Canonicalize memory representation; standardize schemas; apply sensitivity checks. \\

Sunk Cost Fallacy \cite{Kahneman2011} & Agent continues ineffective strategy due to prior investment. Example: agent's memory prioritizes records that justify the past investment in a specialized, high-capacity Massive MIMO antenna (a
gNB node upgrade), causing it to ignore new spectral analysis
showing the node is now underutilized. & Reasoning, Planning, Memory, Verification & Reset historical influence; include diminishing-return detection; flag persistent low-benefit loops. \\

Neglect of Probability / Uncertainty \cite{Tversky1974,Fischhoff1977} & Agent ignores uncertainty, acting on point estimates. Example: Self-healing xApp treats 60\% alarm as certain. & Reasoning, Planning, Verification & Integrate Bayesian or ensemble-based uncertainty; apply confidence thresholds. \\

Status Quo Bias \cite{Samuelson1988,Kahneman2011} & Agent prefers current configuration due to overestimated switching risks. Example: RIC avoids updating cell configuration despite KPIs suggesting optimization. & Planning, Reasoning, Verification & Reduce inertia thresholds; simulate reversible changes. \\

Automation Bias \cite{Kahneman2011} & Over-reliance on automated tool outputs without verification. Example: xApp propagates erroneous rApp fault detection. & Tool Use, Reasoning, Planning, Communication & Mandatory verification; confidence estimation; defer risky actions; share verification status across agents. \\

Survivorship Bias \cite{Carpenter1999,Kahneman2011} & Agent learns only from successes, ignoring failures. Example: a slice-scaling rApp agent trained only on successes ignores failures. & Memory Dataset, Reasoning, Planning & Ensure balanced success/failure sampling; simulate failures; verify performance on complete data. \\
\bottomrule
\end{tabularx}
}
\end{table*}
\vspace{5mm}
\section{Cognitive Biases in LLM-Powered 6G Agents}
\label{sec:biases}

Autonomous decision-making in O-RAN increasingly leverages LLM-powered agents for tasks such as slice orchestration, resource allocation, fault isolation, and multi-agent negotiation. While LLMs provide flexible reasoning and large-context retrieval, they also inherit - and can amplify - well-known cognitive biases. These biases arise from (i) how memories are retrieved and ranked, (ii) how prompts and peer messages condition priors, and (iii) how reward or trust signals are aggregated across episodes. In performance-sensitive settings such as 6G RAN control, even small systematic biases can cascade into large degradations. Table~\ref{tab:bias-taxonomy} enumerates the well-known biases and their emergence in 6G, wherefore we provide a companion exposition that explains each bias in greater depth: its mechanism, a compact mathematical view, a concrete 6G example, likely impacted agentic components, and pragmatic mitigation directions.

\subsection{Overview of Biases, their Impact and Mitigation in 6G}
\rev{
\subsubsection{\textbf{Confirmation Bias}}
An LLM agent exhibits confirmation bias when it selectively retrieves or interprets evidence that supports its existing hypothesis, and thereby under-samples or ignores disconfirming data. Concretely, let $S(\mathcal{D},H)$ denote a selection operator that depends on dataset $\mathcal{D}$ and hypothesis $H$. Then the posterior obtained from the selected data becomes
\begin{equation}
p(H \mid S(\mathcal{D},H)) \;\propto\; p(H)\prod_{x\in S(\mathcal{D},H)} p(x\mid H),
\label{eq:conf_post}
\end{equation}
which biases the posterior toward $H$ by over-weighting supportive likelihood terms. Equivalently, if selection induces a proposal sampling distribution $q(x)\propto \mathbf{1}\{x\in S(\mathcal{D},H)\}p(x)$, an unbiased estimator of an expectation $\mathbb{E}_p[f(X)]$ requires importance weights $w(x)=p(x)/q(x)$; failure to apply such correction yields a biased estimate:
\[
\widehat{\mathbb{E}}[f] = \frac{\sum_{x\sim q} f(x)}{\sum_{x\sim q}1} \quad\neq\quad \mathbb{E}_p[f(X)].
\]
From an information-theoretic perspective, selection skew can be measured by the Kullback--Leibler divergence between the true posterior and the selection-conditioned posterior,
\[
\mathrm{KL}\big(p(H\mid\mathcal{D}) \;||\; p(H\mid S(\mathcal{D},H))\big),
\]
which grows as selection concentrates evidence supporting $H$ \cite{Nickerson1998,Gelman2013}.

In O-RAN, for example, an xApp that diagnoses ``scheduler congestion'' may query only E2 KPIs that corroborate congestion while ignoring interfering telemetry (e.g., interference indicators). Consequently the agent's memory retrieval, reasoning/planning and tool use pipelines each amplify the flawed hypothesis: (i) retrieval will preferentially surface supporting episodes, (ii) reasoning will prematurely converge on a single explanation, and (iii) tool queries and simulations will be framed to validate the hypothesis rather than to falsify it.

\textbf{Mitigation:} To reduce confirmation bias we can combine sampling-correction, explicit falsification, and exploration regularization. In \emph{memory retrieval} enforce \emph{symmetric sampling} across competing hypotheses by drawing samples from both $q_{\text{support}}$ and $q_{\text{refute}}$ with controlled proportions, and apply importance weights to recover unbiased estimates:
\[
\widehat{\mathbb{E}}_p[f] \;=\; \frac{\sum_{x\sim q_{\text{support}}\cup q_{\text{refute}}} w(x) f(x)}{\sum_{x} w(x)}, \qquad w(x)=\frac{p(x)}{q(x)}.
\]
During \emph{reasoning} implement \emph{counterfactual queries} that explicitly compute alternative-likelihood ratios,
\[
\Lambda(x) \;=\; \frac{p(x\mid H_{\text{alt}})}{p(x\mid H_{\text{curr}})},
\]
and preferentially surface cases with large $\Lambda$ to challenge $H_{\text{curr}}$. In \emph{planning} add an exploration regularizer (entropy or uncertainty bonus) to the objective:
\[
\pi^\star = \arg\max_\pi \mathbb{E}_{a\sim\pi}[U(a)] + \beta \mathcal{H}(\pi),
\]
which discourages overly deterministic, belief-consistent plans \cite{sutton2018}. Finally, in \emph{communication} adopt a challenge--response protocol where peer agents must produce disconfirming evidence sampled under an alternative-policy prior before consensus is reached. These interventions reduce the KL shift induced by selective evidence and re-align posterior estimates with the full data-generating process \cite{Nickerson1998, Gelman2013}.

\subsubsection{\textbf{Recency and Primacy Bias}}
Recency (over-weighting recent data) and primacy (over-weighting early data) can be modeled through temporal weighting of observations. A common weighted estimator is
\begin{equation}
\hat{\theta} \;=\; \frac{\sum_{t} w_t x_t}{\sum_t w_t}, \qquad w_t \propto e^{-\lambda |T-t|},
\label{eq:temp_weight}
\end{equation}
where a large $\lambda$ concentrates mass around the most recent ($t\approx T$) or earliest ($t\approx 0$) samples, producing strong recency or primacy effects respectively. Equivalently, an Exponentially Weighted Moving Average (EWMA) obeys the recursive form
\[
\hat{\theta}_{t} = \alpha x_t + (1-\alpha) \hat{\theta}_{t-1},
\]
which is mathematically identical to a first-order Kalman filter under constant observation and process noise assumptions \cite{kalman1960}. Overly large $\alpha$ (equivalently large $\lambda$) yields recency bias, while very small $\alpha$ yields primacy-like inertia.

In practice, fault-management xApps that react to short E2 spikes or that never update thresholds illustrate these failure modes: short-term E2 spikes dominate decisions in the former (recency), while initial thresholds remain dominant in the latter (primacy).

\textbf{Mitigation:} Balance can be achieved by multi-horizon estimators, change-point detection, and temporal confidence bounds. Concretely: in \emph{memory} calibrate decay functions across horizons by maintaining parallel estimators $\{\hat{\theta}^{(h)}\}_{h=1}^H$ with different decay rates $\alpha_h$ and fuse them via weighted averaging,
\[
\hat{\theta}_{\text{fused}}=\sum_{h=1}^H \omega_h \hat{\theta}^{(h)},\qquad \sum_h \omega_h=1.
\]
In \emph{reasoning}, detect change points by monitoring likelihood-ratio statistics or CUSUM statistics and only adapt aggressively when the change is statistically significant \cite{Page1954}. During \emph{planning}, use multi-window averaging or evidence fusion across windows to avoid overreacting to anomalies. Finally, \emph{verification} can enforce temporal confidence bounds (e.g., via bootstrap or Bayesian credible intervals) to prevent reconfigurations driven by transient fluctuations \cite{Gelman2013}.

\begin{figure}[t]
    \centering
    \includegraphics[width=1.0\linewidth]{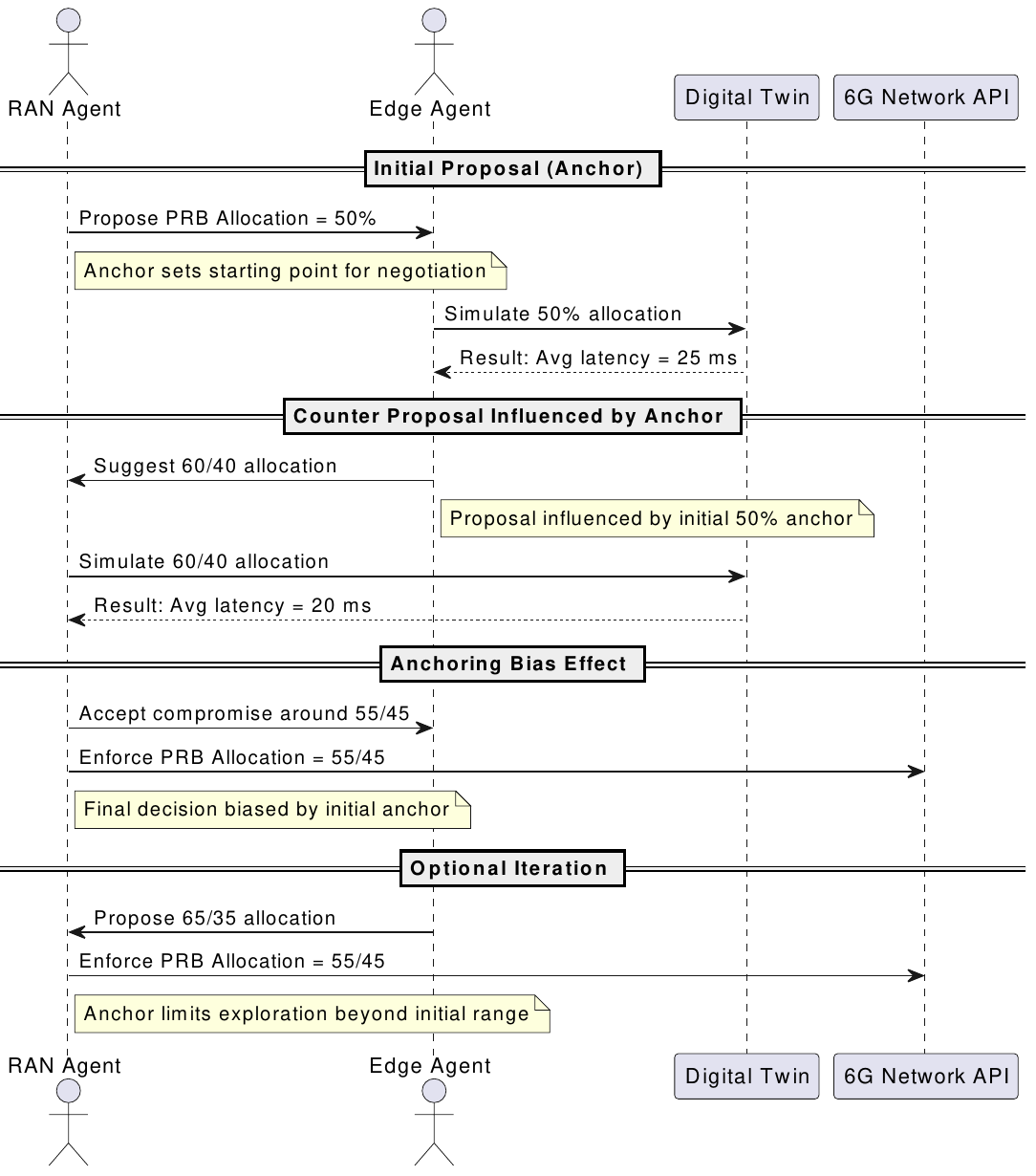}
    \caption{Anchoring bias in 6G agentic negotiation.}
    \label{fig:anchor}
\end{figure}

\subsubsection{\textbf{Anchoring Bias}}
Anchoring occurs when an early piece of information $a_0$ constrains subsequent optimization or negotiation; for instance one can express the anchored decision as a regularized maximization
\begin{equation}
a^\star = \arg\max_a \big(U(a) - \gamma\,d(a,a_0)\big),
\label{eq:anchor_obj}
\end{equation}
where $d(\cdot,\cdot)$ is a deviation penalty and $\gamma$ controls the anchor's strength. This formulation is closely related to regularized empirical risk minimization where an initial estimate acts as a strong prior.

In a resource negotiation example, initial PRB proposals bias the negotiation path (Fig.~\ref{fig:anchor}). Anchoring affects (i) memory/retrieval by biasing which historical outcomes are retrieved (those near $a_0$), (ii) reasoning/planning by limiting search to neighborhoods of $a_0$ even when simulations indicate different optima, and (iii) tool use by restricting DT validation to small perturbations around the anchor.

\textbf{Mitigation:} Reduce anchoring via randomized initialization, decay of anchor influence, and anchor-independent verification. During \emph{planning initialization}, randomize anchor values or sample several anchor candidates $\{a_0^{(k)}\}$ to diversify starting points. In \emph{reasoning} apply a temporal decay to the anchor penalty, e.g.,
\[
\gamma_t = \gamma_0 e^{-\eta t},
\]
so the anchor influence wanes as new evidence accumulates. Within \emph{communication}, refresh shared embeddings or priors to avoid inherited anchoring, and perform \emph{anchor-independent verification} by re-evaluating decisions under perturbed initial conditions to test robustness \cite{Tversky1974}.

\subsubsection{\textbf{Availability Bias}}
Availability bias arises when the agent overestimates the probability of salient or easily retrievable events. If retrieval has a bias function $r(x)\ge0$ that modulates the chance an item $x$ is recalled, the estimated probability of an event $E$ from retrieved samples becomes
\begin{equation}
\hat{p}(E) \;=\; \frac{\sum_x r(x)\mathbf{1}[x\ \text{evidences}\ E]}{\sum_x r(x)}.
\label{eq:avail}
\end{equation}
This estimator is biased away from the true event probability $p(E)$ unless $r(x)\propto 1$ (uniform). For rare-but-memorable alarms, $r(x)$ is large and $\hat p(E)$ overestimates true risk.

Availability bias thus distorts (i) memory and retrieval---favoring high-salience episodes, (ii) reasoning and planning---focusing on easily available scenarios rather than base rates, and (iii) tool use---preferring quick summary tools instead of comprehensive simulations.

\textbf{Mitigation:} Use inverse-probability weighting and stratified sampling. In \emph{memory} apply an inverse-salience reweighting to retrieved items,
\[
w_{\text{inv}}(x) \propto \frac{1}{r(x)+\epsilon},
\]
so that highly salient items are down-weighted during aggregation. During \emph{reasoning}, introduce \emph{salience normalization} via prompts or internal circuitry that enforces base-rate correction (i.e., explicitly combine prior $p(E)$ with retrieved evidence likelihoods). In \emph{planning} and \emph{communication}, enforce \emph{stratified sampling} across event frequency bins to ensure rare but critical events are proportionally represented in deliberation.
\begin{figure}
    \centering
    \includegraphics[width=1.1\linewidth, trim=0.5cm 1cm 0 0, clip]{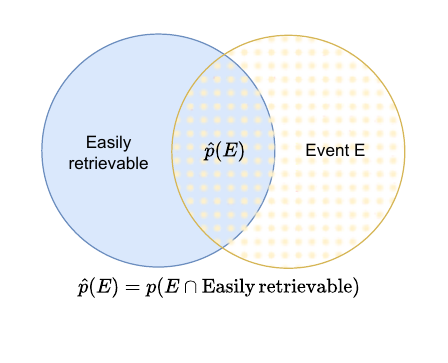}
    \caption{Availability bias concept.}
    \label{fig:avail}
\end{figure}
\subsubsection{\textbf{Authority Bias}}
Authority bias occurs when outputs from a trusted source are over-weighted. A simple linear fusion model is
\begin{equation}
\hat{y} = \tau_s y_s + (1-\tau_s) y_{\text{local}}, \qquad \tau_s \in [0,1],
\label{eq:auth}
\end{equation}
and authority bias corresponds to $\tau_s\gg 0$ even when $y_s$ conflicts with local evidence. More generally, let the trust parameter $\tau_s$ be adaptively updated from observed performance; then a principled update can be
\[
\tau_{s,t+1} = \frac{\exp\big(\alpha\,\mathrm{perf}_{s,t}\big)}{\sum_{s'} \exp\big(\alpha\,\mathrm{perf}_{s',t}\big)},
\]
which is a softmax calibration of source reliabilities based on recent accuracy or calibrated likelihoods \cite{josang2001}.

Authority bias impacts (i) reasoning/planning, where proposals from an ``expert'' agent are accepted without independent validation, and (ii) tool use, where a single authoritative API dominates decisions.

\textbf{Mitigation:} Implement dynamic trust calibration, cross-validation, and redundancy. In \emph{reasoning}, compute $\tau_s$ from an exponentially weighted accuracy track-record and contextual similarity metrics, and decay trust when conflicts with local telemetry appear. During \emph{planning}, perform cross-validation of authoritative recommendations against independent DT simulations or ensemble models (e.g., compare $y_s$ to $\bar y_{\text{ensemble}}$ and reject if $|y_s-\bar y_{\text{ensemble}}|$ exceeds a threshold). In \emph{communication}, require multiple corroborating reports before full acceptance of an authoritative recommendation \cite{josang2001}.

\subsubsection{\textbf{Halo Effect}}
The halo effect occurs when success on one task inflates trust in other tasks. Let $\tau_t^{(j)}$ be the trust in agent (or module) $j$ at time $t$. A simple cross-task update is
\begin{equation}
\tau_{t+1}^{(j)} = \tau_t^{(j)} + \eta\,\mathbf{1}(\text{success}^{(i)}_t)\,\rho_{ij},
\label{eq:halo}
\end{equation}
where $\rho_{ij}$ measures task similarity and $\eta$ is a learning rate. If $\rho_{ij}$ is overestimated the halo effect causes inappropriate cross-domain transfer of confidence (e.g., an xApp that solved energy tuning being over-trusted in latency control).

\textbf{Mitigation:} Constrain cross-domain generalization by maintaining domain-separated latent representations and gating cross-task influence. In \emph{reasoning} maintain separate knowledge partitions or embeddings for each operational domain and only permit transfer when empirical cross-task correlation exceeds a validated statistical threshold. During \emph{planning}, apply a cross-task influence gate $g_{ij}\in[0,1]$ such that effective trust becomes $\tilde\tau^{(j)} = g_{ij}\tau^{(j)}$, and update $g_{ij}$ only when $\rho_{ij}$ is statistically validated (e.g., via hypothesis testing). In \emph{communication}, always include explicit domain context and task-specific performance metrics to prevent blind transfer of authority.

\begin{figure}[t]
    \centering
    \includegraphics[width=1.1\linewidth, trim=0.5cm 0 0 0, clip]{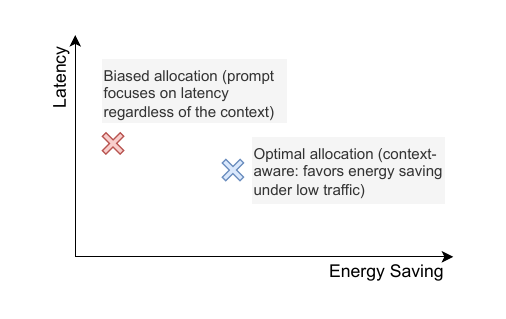}
    \vspace{-1cm}
    \caption{Suggestion Bias shifts the agent's decision in the multi-objective space.}
    \label{fig:sugg}
\end{figure}

\subsubsection{\textbf{Suggestion / Prompting Bias}}
Prompt wording shifts the model's logits by an additive or multiplicative bias. A first-order model is
\begin{equation}
\text{logits} \leftarrow \text{logits} + \beta\, s(\text{prompt}),
\label{eq:prompt}
\end{equation}
where $s(\cdot)$ encodes prompt-derived bias and $\beta$ scales its strength. For example, the prompt ``latency is the primary goal'' corresponds to a score vector $s$ that elevates latency-optimizing actions, thereby steering the agent toward those actions even when they may not be globally optimal.

This affects (i) memory/retrieval by seeding queries with leading phrases that bias fetched records, (ii) reasoning/planning by introducing prior assumptions into chain-of-thought, and (iii) tool use by constraining DT simulation ranges.

\textbf{Mitigation:} Normalize and track provenance of prompts, detect distributional shifts, and perform controlled re-prompting. In \emph{communication}, canonicalize prompts to neutral forms before downstream processing. In \emph{memory}, store prompt provenance metadata and include provenance-aware de-biasing when scoring retrieved items. In \emph{reasoning}, monitor prompt-embedding statistics (e.g., detect mean shifts) and trigger controlled re-evaluation under semantically equivalent but neutrally phrased prompts. Additionally, run DT validation under broadened parameter ranges rather than the narrow ranges implied by the original prompt to guard against framing-induced narrow search.
\begin{figure}[t]
    \centering
    \includegraphics[width=1.0\linewidth]{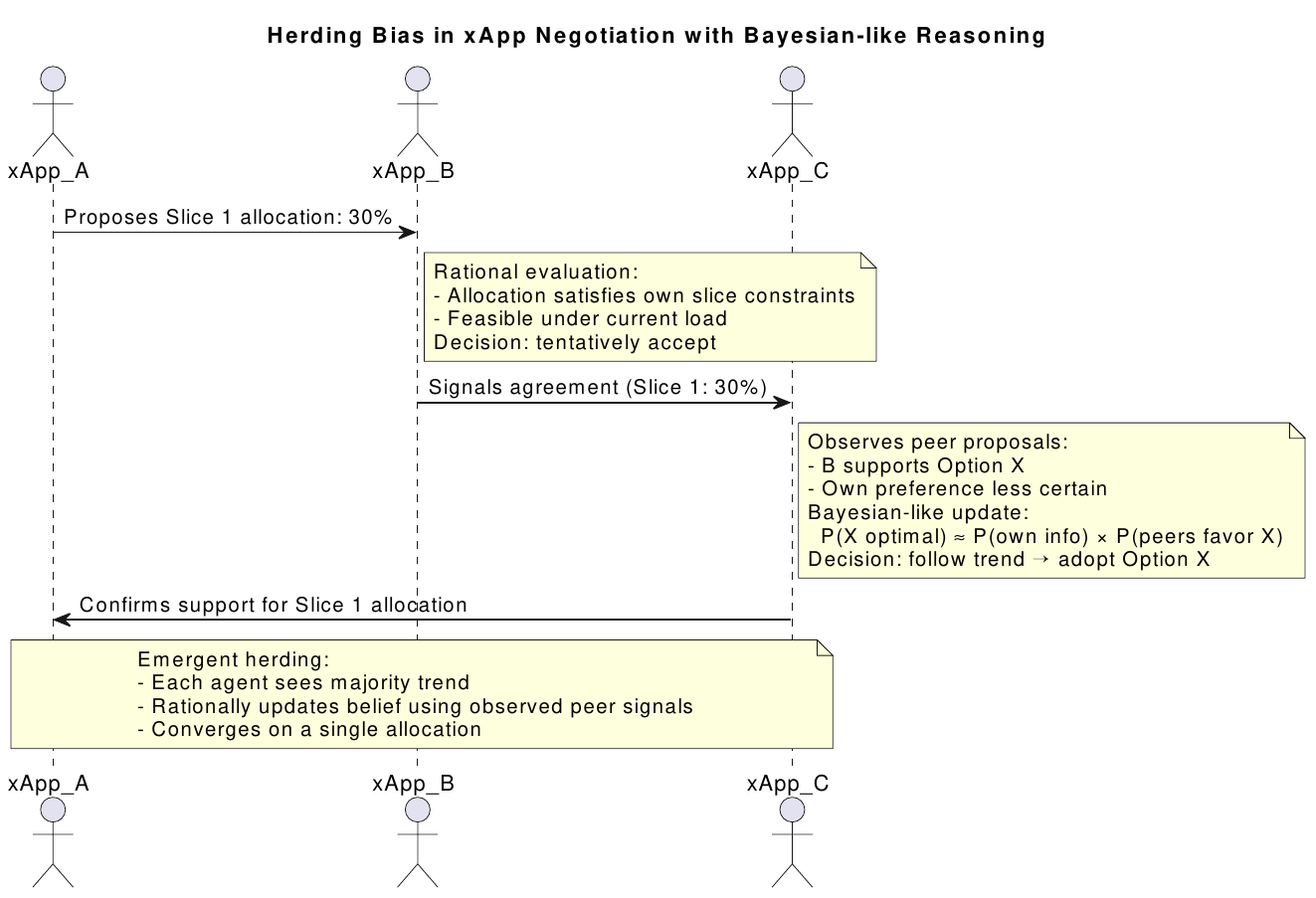}
    \caption{Groupthink emergence.}
    \label{fig:group}
\end{figure}
\vspace{5mm}
\subsubsection{\textbf{Groupthink / Herding}}
Groupthink or herding arises when conformity to peer agents constrains independent reasoning. In a consensus-regularized multi-agent objective:
\begin{equation}
\min_{a_i}\sum_i \mathcal{L}_i(a_i) + \lambda \sum_i \|a_i-\bar{a}\|^2,
\label{eq:herd_obj}
\end{equation}
a high $\lambda$ forces rapid consensus. Consider an agent C that is uncertain: after observing peers favoring option $X$, C performs a Bayesian-like combination of its own signal and the social signal. In a simple approximation,
\begin{equation}
\Pr(X_{\text{optimal}} \mid \text{data},\ \text{peers}) \propto \Pr(\text{own info})\times \Pr(\text{peers favor }X),
\label{eq:bayes}
\end{equation}
which illustrates how weak private evidence can be overwhelmed by strong social evidence \cite{bikhchandani1992}. This social reinforcement reduces diversity and can lead to brittle consensus.

\textbf{Mitigation:} Preserve diversity via independent rollouts, ensemble disagreement, dissent propagation, and dynamic consensus relaxation. In \emph{planning}, require independent rollouts of candidate strategies before synchronization, and in \emph{reasoning} introduce an \emph{ensemble disagreement} objective that rewards hypothesis diversity, e.g.,
\[
\min_{\theta_1,\dots,\theta_M} \sum_{m} \mathcal{L}(\theta_m) - \gamma \sum_{m\neq m'} \mathrm{D}(\theta_m,\theta_{m'}),
\]
where $\mathrm{D}$ is a diversity-promoting functional. In \emph{communication}, adopt dissent-propagation protocols to surface minority viewpoints, and dynamically reduce the consensus regularizer $\lambda_t$ over deliberation rounds:
\[
\lambda_{t+1} = \lambda_t \cdot \delta,\qquad \delta\in(0,1),
\]
so conformity pressure relaxes as more evidence accumulates.

\bigskip

\noindent\textbf{Remarks on combining mitigations.} Many of the mitigation strategies above share common building blocks from statistical learning and decision theory: importance weighting (for selection bias), multi-horizon fusion and change-point detection (for temporal bias), entropy/uncertainty regularization (for exploration vs.\ exploitation), cross-validation and softmax-based trust calibration (for authority/halo), and ensemble/diversity objectives (for groupthink). Where appropriate, these mechanisms should be deployed together in the agentic pipeline (memory, reasoning, planning, tool use, and communication) so that one module's counter-biasing does not inadvertently create another bias downstream. The short bibliography below lists accessible references for the learning-theoretic tools and the behavioural origins of the biases.
}

\begin{table*}[t]
\centering
\caption{\rev{Theoretical and Learning-Theoretic Characterization of Cognitive Biases in LLM-Based 6G Agents}}
\label{tab:theory_biases}
\renewcommand{\arraystretch}{1.2}
\begin{tabularx}{\textwidth}{l X X X X}
\hline
\textbf{Bias} &
\textbf{Learning-Theoretic Origin} &
\textbf{Mathematical Abstraction} &
\textbf{Primary Computational Mechanism} &
\textbf{Related Theory and References} \\
\hline

Confirmation &
Biased posterior updating and selective sampling &
$p(H|S(\mathcal D,H)) \propto p(H)\prod_{x\in S}p(x|H)$ &
Hypothesis-consistent evidence filtering &
Bayesian inference, PAC-Bayes \cite{Nickerson1998,Gelman2013,mcallester1999} \\

Recency / Primacy &
Temporal discounting and non-stationary estimation &
$\hat{\theta}=\frac{\sum_t w_tx_t}{\sum_t w_t}$ &
Exponential decay of historical samples &
Reinforcement learning, adaptive filtering \cite{sutton2018,kalman1960} \\

Anchoring &
Regularized utility optimization with priors &
$\arg\max_a U(a)-\gamma d(a,a_0)$ &
Strong prior anchoring in objective landscape &
Bayesian priors, constrained optimization \cite{Tversky1974,Gelman2013} \\

Availability &
Salience-weighted sampling and retrieval bias &
$\hat p(E)=\frac{\sum_x r(x)\mathbf{1}_E}{\sum_x r(x)}$ &
Over-representation of vivid samples &
Sampling bias, heuristic inference \cite{Tversky1974} \\

Authority &
Source-weighted decision fusion &
$\hat y=\tau_sy_s+(1-\tau_s)y_l$ &
Static trust over-weighting &
Trust modeling, opinion fusion \cite{josang2001,Gelman2013} \\

Halo Effect &
Cross-task confidence propagation &
$\tau_{t+1}^{(j)}=\tau_t^{(j)}+\eta\rho_{ij}$ &
Unconstrained transfer learning &
Multi-task learning, transfer bias \cite{sutton2018} \\

Suggestion / Prompting &
Logit-space perturbation and conditioning &
$\text{logits}\leftarrow \text{logits}+\beta s$ &
Prompt-induced activation shifts &
Representation learning, prompting theory \\

Groupthink / Herding &
Consensus regularization and social learning &
$\min\sum_i\mathcal L_i+\lambda\|a_i-\bar a\|^2$ &
Social signal amplification &
Informational cascades, game theory \cite{bikhchandani1992} \\

Framing Effect &
Utility distortion under representation changes &
$a(\phi(x))\neq a(\psi(x))$ &
Context-dependent utility mapping &
Prospect theory, invariance learning \cite{Tversky1974} \\

Sunk Cost &
Path-dependent utility accumulation &
$U=\mathbb E[B]-C+\alpha f(S)$ &
Irrecoverable-cost reinforcement &
Dynamic programming, behavioral economics \cite{Tversky1974} \\

Neglect of Uncertainty &
Point-estimate optimization &
$\arg\max_a U(a,\mathbb E[x])$ &
Variance suppression &
Bayesian learning, ensemble methods \cite{Gelman2013} \\

Status Quo &
Switching-cost regularization &
$\Delta U>c_{\text{switch}}$ &
Inertia-dominated optimization &
Decision theory, risk aversion \cite{Tversky1974} \\

Automation &
Tool-output dominance &
$a=f(\text{tool})$ &
Verification bypass &
Human-in-the-loop systems \\

Survivorship &
Selection-truncated learning &
$\mathbb E[\hat\theta|S=1]\neq\theta$ &
Failure-data elimination &
Selection bias, empirical risk \cite{auer2002} \\

\hline
\end{tabularx}
\end{table*}

\begin{figure}[t]
    \centering
    \includegraphics[width=1.0\linewidth, trim=0 0 0 0, clip]{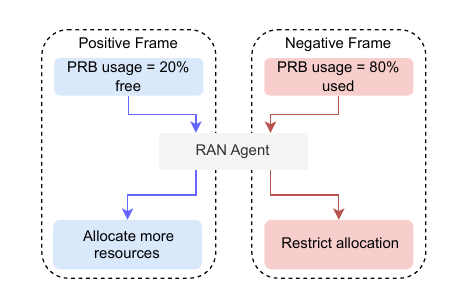}
    \vspace{-7mm}
    \caption{Framing effect concept.}
    \label{fig:fram}
\end{figure}

\rev{\subsubsection{\textbf{Framing Effect}}
Framing occurs when semantically equivalent descriptions of the same underlying state lead to different decisions, and thus the agent's action depends on the presentation map (or frame). Formally, let $\phi,\psi$ be two framing maps with $\phi(x)\!\sim\!\psi(x)$ (i.e., they convey the same informational content up to re-encoding), yet the agent's policy $a(\cdot)$ satisfies
\[
a(\phi(x)) \neq a(\psi(x)).
\]
To ground this behavior in decision theory, consider a prospect-theory style value function $v(\cdot)$ and a frame-dependent reference point $r_{\phi}$ induced by frame $\phi$. Then the framed decision can be written as the solution of a framed utility maximization
\begin{equation}
a^\star_\phi \;=\; \arg\max_a\ \mathbb{E}\big[ v\big(R(a)-r_{\phi}\big)\big],
\label{eq:frame_decision}
\end{equation}
where $R(a)$ is the (random) reward of action $a$. In prospect-theory parameterizations one commonly uses
\[
v(u) = \begin{cases}
u^{\alpha}, & u\ge 0,\\[4pt]
- \lambda (-u)^{\beta}, & u<0,
\end{cases}
\qquad \alpha,\beta\in(0,1),\ \lambda>1,
\]
so that changing the reference $r_\phi$ (a gain vs.\ loss framing) systematically alters the sign and curvature of the argument to $v(\cdot)$ and hence the optimizer in (\ref{eq:frame_decision}) \cite{Tversky1981,Kahneman2011}. Equivalently, framing can be captured as a change in the agent's internal prior or loss function: $\ell_\phi(a)\neq\ell_\psi(a)$, producing different minimizers $\arg\min_a \mathbb{E}[\ell_\phi(a\!;\!x)]$ and $\arg\min_a \mathbb{E}[\ell_\psi(a\!;\!x)]$ even when $\phi(x)\sim\psi(x)$.

For example, PRB usage described as ``20\% free'' versus ``80\% used'' corresponds to different reference points $r_{\phi}$ and $r_{\psi}$ and therefore to different framed utilities in (\ref{eq:frame_decision}), which changes the allocation decision (cf.\ Fig.~\ref{fig:fram}). Consequently, framing influences (i) memory/retrieval, because prompts phrased to emphasize gains preferentially surface gain-framed records; (ii) reasoning/planning, because loss-framed descriptions induce more conservative, risk-averse plans; and (iii) tool use, because labeled metrics (``high utilization'' vs.\ ``efficient usage'') guide downstream validation and reconfiguration.

\textbf{Mitigation:} To counter framing bias we should (i) canonicalize internal representations so that semantically equivalent inputs map to the same latent encoding, i.e.\ enforce $\mathcal{E}(\phi(x))\approx\mathcal{E}(\psi(x))$, (ii) train for paraphrase- and reference-invariance by augmenting training data with alternative framings and enforcing stability of the decision rule (e.g., add a training penalty $\|\pi(\phi(x))-\pi(\psi(x))\|^2$), and (iii) implement a verification layer that performs sensitivity checks by re-evaluating decisions under systematically rephrased inputs and by comparing framed optima $a^\star_\phi,a^\star_\psi$. These steps create explicit invariances in both representation and objective, thereby reducing frame-driven divergences \cite{Tversky1981,Kahneman2011}.


\subsubsection{\textbf{Sunk Cost Fallacy}}
The sunk cost fallacy arises when past, irrecoverable investments $S$ improperly affect present choices. While rational decision theory dictates that sunk costs should not influence future-optimal actions, biased agents internalize an additive utility term dependent on $S$:
\begin{equation}
U_{\text{biased}}(a) \;=\; \mathbb{E}[B(a)] - C(a) + \alpha\, f(S),
\label{eq:sunk_simple}
\end{equation}
with $\alpha>0$ and $f(\cdot)$ an increasing function (e.g., $f(S)=\log(1+S)$). In an MDP formalism, the correct Bellman equation for the value of state $s$ is
\[
V^\star(s) \;=\; \max_a\ \big[ r(s,a) + \gamma \mathbb{E}_{s'|s,a} V^\star(s')\big],
\]
and it should ignore costs that do not affect future transitions or rewards. However, if the agent augments instantaneous reward $r(s,a)$ with a sunk-cost term depending on historical investment $S$ (so that $r_{\text{biased}}(s,a)=r(s,a)+\alpha f(S)$), then its learned policy $\pi_{\text{biased}}$ solves a different dynamic program and therefore can be strictly suboptimal relative to the true optimal policy $\pi^\star$ that correctly treats $S$ as irrelevant to future payoffs \cite{ARKES1985124,Samuelson1988}.

Practically, this mechanism explains why an agent might continue optimizing a massive MIMO gNB node---because $S$ (the expensive upgrade) inflates the perceived utility of actions that defend that investment---despite spectral evidence suggesting reallocation to backhaul optimization would be more efficient. Thus sunk cost affects (i) memory/retrieval by prioritizing supporting records, (ii) reasoning/planning by biasing optimization loops, and (iii) tool use by focusing simulations on marginal improvements around the sunk asset.

\textbf{Mitigation:} Countermeasures include (i) periodically {\em resetting historical influence} by projecting state representations to remove or discount historical-investment components (e.g., replace $S$ with $\tilde S=\lambda_S S$ with $\lambda_S\downarrow 0$ over time), (ii) including diminishing-return detectors that monitor marginal gains $\Delta B/\Delta \text{effort}$ and trigger reallocation when these fall below thresholds, and (iii) tagging irrecoverable expenditures in memory so they are treated as annotated history rather than continuing objectives. Importantly, treating $S$ as an exogenous label (and not as part of the reward) restores correct Bellman updates and prevents the agent's dynamic program from being misspecified \cite{ARKES1985124,Kahneman2011}.


\begin{table*}[t]
\centering
\caption{\rev{Operational Impact of Cognitive Biases in LLM-Driven 6G Agentic Systems}}
\label{tab:system_biases}
\renewcommand{\arraystretch}{1.2}
\begin{tabularx}{\textwidth}{l X X X X}
\hline
\textbf{Bias} &
\textbf{Affected Pipeline Stages} &
\textbf{Typical 6G Scenario} &
\textbf{Primary Failure Mode} &
\textbf{Key Mitigation Strategies} \\
\hline

Confirmation &
Memory, Reasoning, Tool Use &
Selective KPI querying in congestion diagnosis &
Persistent misdiagnosis &
Symmetric retrieval, counterfactual reasoning \\

Recency / Primacy &
Memory, Planning, Tools &
Overreaction to short-term spikes &
Unstable configurations &
Multi-horizon averaging, change-point detection \\

Anchoring &
Reasoning, Planning, Communication &
PRB negotiation initialization &
Narrow bargaining range &
Anchor randomization, decay weighting \\

Availability &
Memory, Planning, Tools &
Alarm log overreaction &
Poor rare-event preparedness &
Inverse-salience weighting, stratified sampling \\

Authority &
Reasoning, Tools, Communication &
Vendor rApp dominance &
Unverified acceptance &
Dynamic trust calibration, cross-validation \\

Halo Effect &
Reasoning, Tools &
Over-trust of successful xApp &
Cross-domain misconfiguration &
Domain separation, influence gating \\

Suggestion / Prompting &
Memory, Reasoning, Tools &
Directive prompt bias &
Predetermined optimization &
Prompt normalization, re-prompting \\

Groupthink / Herding &
Memory, Planning, Coordination &
Slice allocation consensus &
Premature convergence &
Independent rollouts, dissent propagation \\

Framing Effect &
Memory, Reasoning, Tools &
Gain/loss KPI presentation &
Risk miscalibration &
Canonicalization, paraphrase invariance \\

Sunk Cost &
Memory, Planning, Tools &
Persistent gNB optimization &
Resource lock-in &
Historical discounting, loop detection \\

Neglect of Uncertainty &
Memory, Reasoning, Tools &
Deterministic alarm response &
Brittle allocations &
Bayesian inference, uncertainty exchange \\

Status Quo &
Planning, Tools &
Static spectrum slicing &
Opportunity loss &
Failure modeling, cost quantification \\

Automation &
Reasoning, Verification, Communication &
Fault-triggered shutdown &
Cascading failures &
Mandatory verification, confidence tracking \\

Survivorship &
Memory, Reasoning, Planning &
Success-only training data &
Over-optimistic policies &
Failure logging, inverse weighting \\

\hline
\end{tabularx}
\end{table*}


\begin{figure}
    \centering
    \includegraphics[width=1.0\linewidth, trim=0cm 2cm 0 0, clip]{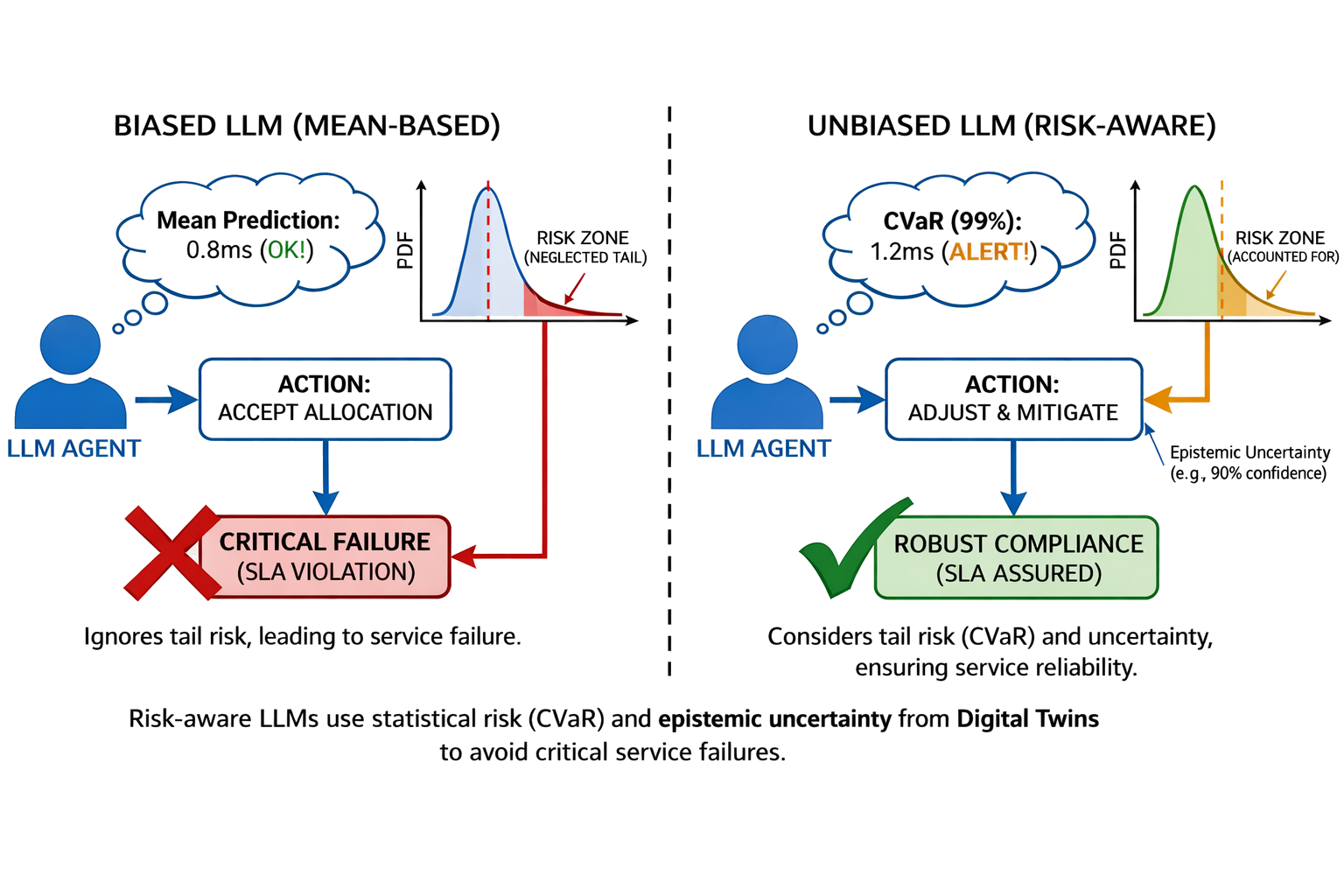}
    \caption{\rev{Uncertainty negelect by reasoning over the mean instead of the tail.}}
    \label{fig:uncert}
\end{figure}
\subsubsection{\textbf{Neglect of Uncertainty}}
Neglect of uncertainty occurs when an agent reduces rich probabilistic beliefs to point estimates and then optimizes as though those estimates were certain. Concretely, the biased decision rule often takes the form
\begin{equation}
   a_{\text{biased}} \;=\; \arg\max_a\ U\big(a,\ \mathbb{E}[x]\big), 
\end{equation}

whereas the Bayes-optimal decision minimizes Bayes risk
\begin{equation}
\delta^\star \;=\; \arg\min_\delta\ \mathbb{E}_{\theta}\mathbb{E}_{x\mid\theta}\big[ L(\theta,\delta(x))\big],
\label{eq:bayes_risk}
\end{equation}
thereby integrating model uncertainty into action selection \cite{berger1985,Gelman2013}. Ignoring uncertainty is equivalent to replacing the posterior $p(\theta\mid \mathcal{D})$ by a delta at a point estimate $\hat\theta$, which in general increases expected loss:
\begin{equation}
  \mathbb{E}_{\theta}\mathbb{E}_{x\mid\theta}L(\theta,\delta_{\hat\theta}(x)) \;\ge\; \mathbb{E}_{\theta}\mathbb{E}_{x\mid\theta}L(\theta,\delta^\star(x)).  
\end{equation}

From a practical standpoint, neglect manifests when a self-healing xApp treats a 60\% confidence alarm as if it were certain and thus executes an all-or-nothing action. This bias corrupts (i) memory/retrieval by storing point estimates without variances, (ii) reasoning/planning by optimizing for expected values without risk buffers, and (iii) tool use by ignoring predictive error bars.

\textbf{Mitigation:} Remedies include (i) integrating Bayesian or ensemble methods to propagate posterior uncertainty through downstream decision modules (e.g., use posterior predictive distributions in (\ref{eq:bayes_risk})), (ii) employing risk-aware objectives such as Conditional Value-at-Risk (CVaR) or robust max-min criteria
\begin{equation}
 \max_a \ \min_{q\in\mathcal{Q}} \mathbb{E}_{\theta\sim q}[U(a,\theta)],   
\end{equation}

to protect against tail risk, and (iii) enforcing confidence thresholds before actioning critical changes. Implementation can rely on approximate Bayesian techniques (dropout-as-Bayesian \cite{gal2016}) or explicit ensemble variance estimation \cite{bishop2006, Gelman2013}.


\subsubsection{\textbf{Status Quo Bias}}
Status quo bias reflects an aversion to changes whose immediate cost (or perceived risk) outweighs the expected gain. A simple threshold condition captures this:
\begin{equation}
   \Delta U \;=\; U_{\text{new}} - U_{\text{current}} \;>\; c_{\text{switch}}, 
\end{equation}

where $c_{\text{switch}}$ is an (often overestimated) switching cost. In a sequential decision model with discounting, adding a switching penalty to the reward yields
\begin{equation}
  r_{\text{biased}}(s,a,s') = r(s,a) - \mathbf{1}\{a\neq a_{\text{current}}\}\,c_{\text{switch}},  
\end{equation}

which produces inertia: the optimal policy will prefer a continuation action unless the long-run gain outweighs $c_{\text{switch}}$. Thus, even when a re-slice (e.g., 60/40) yields a long-term 15\% efficiency improvement as predicted by a DT, the agent may prefer incremental 1\% tweaks if it overestimates short-term disruption.

\textbf{Mitigation:} To overcome status quo bias, (i) make opportunity costs explicit by augmenting tool outputs with quantitative estimates of lost gain from inaction, (ii) simulate failure modes for the current configuration to expose hidden risks, and (iii) include planned, reversible change experiments (A/B rollouts) that reduce perceived switching costs and provide empirical evidence about transient losses versus long-term gains \cite{Samuelson1988}.


\subsubsection{\textbf{Automation Bias}}
Automation bias is the undue deference to automated outputs in place of independent verification. Let the automated tool produce recommendation $y$ with reliability parameter $\tau\equiv\Pr(y\ \text{correct})$. Suppose verification costs $c_v$ and that verifying reduces expected loss by $\Delta L$. A rational agent verifies iff $\Delta L > c_v$. However, an automation-biased agent skips verification, effectively following
\begin{equation}
    a = f(y)
\end{equation}

instead of the verification-aware rule
\begin{equation}
a = f\big(y,\ \text{verify}(\cdot)\big),
\end{equation}
and hence it incurs additional expected loss when $\tau$ is low or nonstationary \cite{parasuraman2000}.

In networks, this explains immediate, unverified cell shutdowns following a faulty rApp alarm. The bias therefore degrades (i) reasoning/planning by short-circuiting independent checks, (ii) tool use/verification by bypassing corroboration, and (iii) communication by propagating unverified signals.

\textbf{Mitigation:} Introduce mandatory verification gates for high-impact actions, propagate verification status and confidence scores with every automated output, and compute expected-value-of-information for verification so that the agent makes explicit, cost-sensitive checks (verify iff $\Delta L>c_v$). In multi-agent settings, require corroboration from independent sensors or models before committing to irreversible actions \cite{parasuraman2000}.

\begin{figure}
    \centering
    \includegraphics[width=1.0\linewidth]{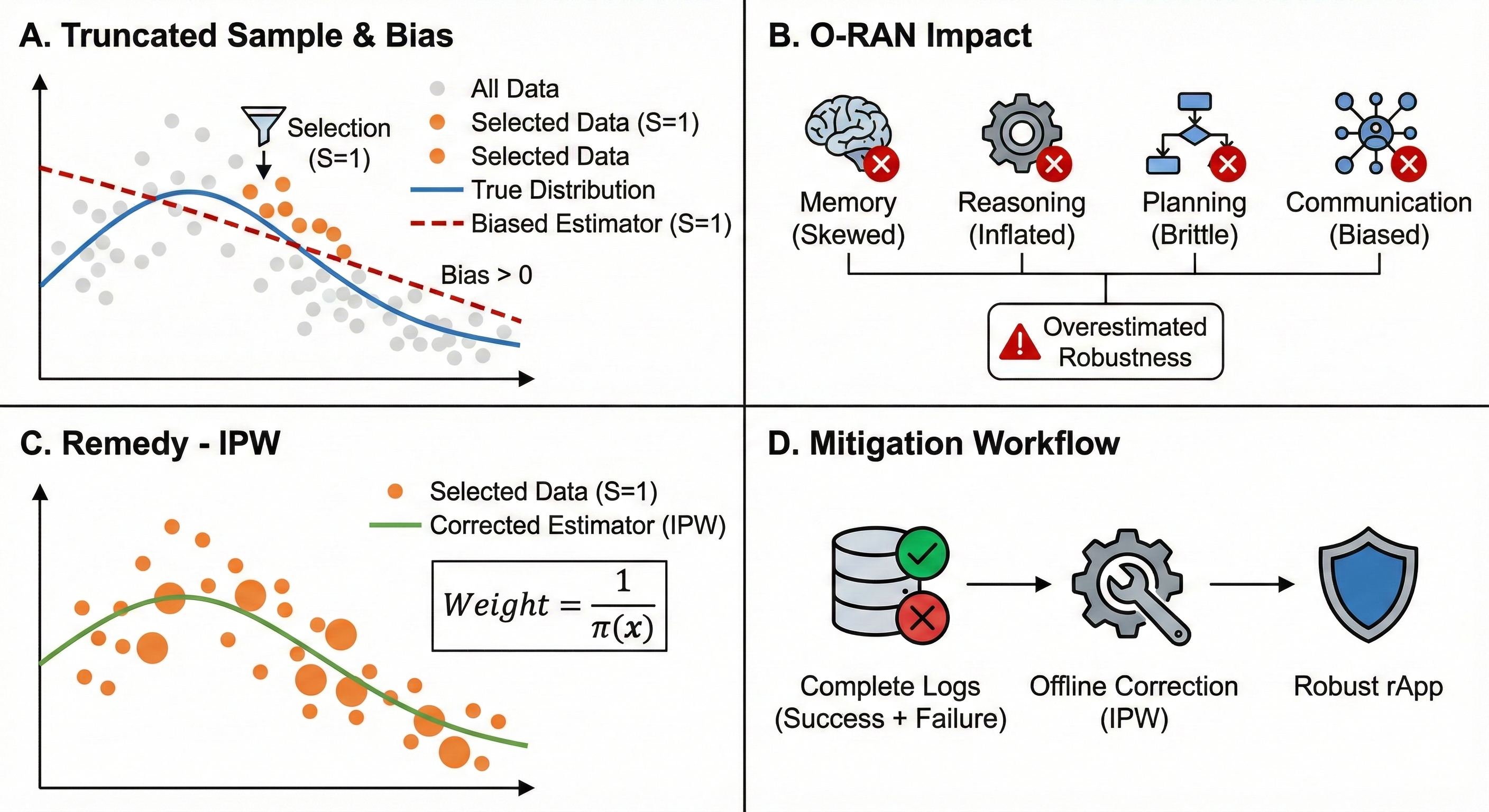}
    \caption{\rev{Concept and mitigation of the survivorship bias in telecom.}}
    \label{fig:surv}
\end{figure}
\subsubsection{\textbf{Survivorship Bias}}
Survivorship bias appears when learning uses only successful (surviving) instances $S=1$, which truncates the sample and biases estimators. Formally, if $\hat\theta$ is estimated from the conditional distribution $p(x\mid S=1)$, then
\begin{equation}
  \mathbb{E}[\hat{\theta}\mid S=1] \;-\; \theta_{\text{true}} \;\neq\; 0,  
\end{equation}

whenever the selection mechanism depends on unobserved variables correlated with the target. This is a classical selection problem; Heckman's correction and inverse-probability weighting (IPW) are standard remedies. In particular, an unbiased estimator can be recovered by reweighting samples with the inverse selection probability:
\begin{equation}
   \widehat{\mathbb{E}}_{ \text{corrected}}[f(X)] \;=\; \frac{\sum_{i: S_i=1} \frac{1}{\pi(X_i)} f(X_i)}{\sum_{i: S_i=1} \frac{1}{\pi(X_i)}}, 
\end{equation}

where $\pi(x)=\Pr(S=1\mid x)$ \cite{heckman1979,rosenbaum2014}.

In the O-RAN context, a slice-scaling rApp trained only on deployed (successful) configurations will ignore failure modes and thus overestimate robustness. Therefore survivorship bias impacts (i) memory by selective storage of winners, (ii) reasoning by inflating success probabilities, (iii) planning by preferring brittle strategies, and (iv) communication by propagating skewed summaries.

\textbf{Mitigation:} Maintain complete success/failure logs, apply inverse-truncation weighting or Heckman-style selection corrections during offline learning, simulate failures in planning to probe robustness, and ensure training/evaluation datasets intentionally include negative examples and decommissioned-element diagnostics \cite{heckman1979,rosenbaum2014}.

}

\begin{figure*}[t]
    \centering
    \includegraphics[width=0.75\linewidth, trim=0.5cm 0 0 0, clip]{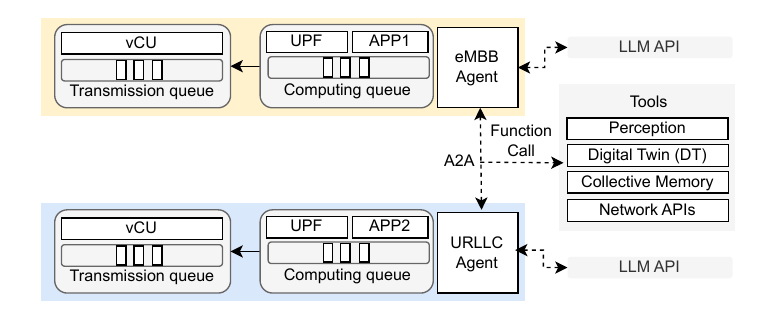}
    \caption{Use-case 1 setup.}
    \label{fig:uc1}
\end{figure*}
\subsection{System-Level Bias Mitigation in 6G Architectures}
Beyond representational and agent-level mitigation detailed before, achieving robust and bias-resilient autonomy in 6G networks requires system-level mitigation mechanisms that operate across orchestration layers. One promising approach is the integration of \emph{adversarial evaluation loops}, where agents are periodically challenged with worst-case or counterfactual scenarios designed to expose latent biases in their decision policies. These loops can detect emergent herding, overfitting to framing effects, providing feedback for reasoning and planning. System-level mitigation also includes ensemble-based orchestration, in which multiple independently trained reasoning agents contribute to decisions, with their outputs aggregated to reduce the impact of individual biases. Within 6G O-RAN, these techniques can be embedded into the Service Management and Orchestration (SMO) layer, allowing xApps and rApps to participate in continuous adversarial evaluation cycles, log bias-relevant metrics, and adjust internal priors dynamically. The overarching principle is that bias-awareness must extend from local agent cognition to network-wide orchestration, ensuring that both the emergent behaviors of distributed agents and the global system policies remain robust against subtle cognitive and systemic biases.

\rev{

\subsection{Discussion on the Costs of Bias-Mitigation in Agentic Systems}

The transition from heuristic-driven Large Language Model (LLM) agents to formally mitigated agentic systems introduces a \emph{mitigation tax} that must be weighed against the gains in reliability. While the proposed strategies effectively align posterior estimates with the true data-generating process, they impose non-trivial burdens across several dimensions. The incorporation of these methods introduces four primary categories of cost, namely,
i) \textbf{Inference and Token Consumption:} To counter confirmation bias, replacing greedy retrieval with symmetric sampling (Eq. 6) across $q_{support}$ and $q_{refute}$ essentially doubles the input context and retrieval operations. Furthermore, computing counterfactual likelihood ratios $\Lambda(x)$ (Eq. 7) necessitates additional forward passes to evaluate alternative-likelihoods, leading to a significant increase in total tokens per decision cycle.
ii) \textbf{Structural Complexity and Statefulness:} Mitigating temporal biases through multi-horizon estimators (Eq. 11) requires the agent's architecture to maintain $H$ parallel versions of the network state. This transforms a stateless agent into a high-overhead stateful system that must continuously track cumulative sum (CUSUM) statistics and temporal confidence bounds, placing a heavy burden on edge-device memory and storage.
iii) \textbf{Convergence Latency and Agility:} The introduction of an exploration regularizer $\beta H(\pi)$ (Eq. 8) into the planning objective intentionally discourages deterministic behavior. While this prevents anchoring, it inherently slows down policy convergence, potentially introducing "decision inertia" in scenarios requiring sub-millisecond reactions, such as rapid beam-steering or fault recovery.
iv) \textbf{Communication and Protocol Overhead:} Challenge-response protocols for multi-agent negotiation require peer agents to generate disconfirming evidence before reaching consensus. This expands the communication topology from a simple broadcast to an iterative, multi-turn dialogue, significantly increasing the signaling load on the 6G management plane.

\begin{table}[htbp]
\centering
\caption{\rev{Operational Costs of Bias-Mitigation}}
\label{tab:mitigation_costs}
\renewcommand{\arraystretch}{1.3}
\footnotesize 
\begin{tabularx}{\columnwidth}{|X|X|X|}
\hline
\textbf{Bias / Mitigation} & \textbf{Resource Impact} & \textbf{Perf. Trade-off} \\ \hline
\textbf{Confirmation} \newline Importance Sampling & High GPU Compute & Latency increase \\ \hline
\textbf{Recency/Primacy} \newline Multi-horizon $\hat{\theta}^{(h)}$ & Memory Footprint & State Complexity \\ \hline
\textbf{Anchoring} \newline Reg. $H(\pi)$ & Iterative Planning & Slower Convergence \\ \hline
\textbf{Groupthink} \newline Challenge-Response & Signaling Load & Bandwidth Spike \\ \hline
\textbf{Framing} \newline Canonicalization & CPU Overhead & System Rigidity \\ \hline
\textbf{Uncertainty} \newline Bayesian Priors & Logic Overhead & Conservative Action \\ \hline
\end{tabularx}
\end{table}

In summary, the cost of fairness in 6G agentic systems is fundamentally a trade-off against throughput. A "biased" agent may reach a decision in a single inference pass, whereas a "mitigated" agent requires a sophisticated ensemble of statistical checks and multi-turn reasoning. However, in critical infrastructure, the potential catastrophic cost of a biased decision (e.g., localized network blackouts or massive resource underutilization) justifies the incremental investment in these robust computational safeguards.

\subsection{General Note on Baselines}
As noted by Xi et al. \cite{risepotentiallargelanguage}, comparing LLM agents to specialized DRL baselines often results in a category error because LLMs are designed for generalized reasoning and complex tool-use in open-ended environments, whereas DRL excels in narrow, high-frequency control loops. Therefore, to ensure a fair and insightful evaluation, our comparisons focus on isolating the impact of cognitive bias mitigation within the LLM architecture itself.

}

\section{Use Case 1: Mitigating Anchoring Bias in Network Slicing Agentic Negotiation}

\subsection{Setup}

\revi{We study the mitigation of anchoring bias in a 6G system by focusing on the dynamic allocation of the shared RAN bandwidth (60 MHz) between the eMBB and URLLC network slices\footnote{\rev{The source code of this use case is available for non commercial use at: https://github.com/HatimChergui/agentic-negotiation-anchoring-bias-mitigation}} as depicted in Figure \ref{fig:uc1}. The system architecture employs separate, autonomous LLM-powered agents, each representing a slice's resource manager, optimizing distinct objectives under strict SLA tail-latency requirements (50 ms for eMBB and 10 ms for URLLC). 

To solve the critical issue of high response times traditionally incurred when using distant commercial LLMs (e.g., Gemini) for real-time control, the agents in this framework are powered by a locally hosted Small Language Model (SLM), specifically the lightweight \texttt{otel-llm-1b-it} model. This ensures sub-second inference latencies (0.95s mean), making the negotiation protocol fully compatible with the operational timescales of the O-RAN non-Real-Time RAN Intelligent Controller (non-RT RIC). 

The agents operate in an iterative negotiation framework, following a three-stage loop within each trial: i) \emph{Digital Twin (DT) consultation,} where agents synchronize their internal Digital Twins to the real network state. The DT uses a fluid-flow queuing model, rigorously evaluated via Conditional Value at Risk (CVaR), to predict the strict tail-latency and energy consumption for any candidate bandwidth proposal based on realistic O-RAN traffic bursts \cite{rezazadeh2022specialization}; ii) \emph{Memory consultation:} Agents consult a shared vanilla memory of past successful agreements to inform their negotiation strategy; and iii) \emph{Negotiation protocol,} where agents engage in a fixed-round, turn-based protocol utilizing PID-like proportional step sizes to reach a feasible joint bandwidth agreement where both agents' utilities exceed a strict acceptance threshold. 

\subsection{Scenarios}

To specifically isolate and evaluate the effect of anchoring bias, we compare two primary initial proposal (anchor) strategies, both utilizing the same underlying SLM agents and vanilla Memory:

\begin{itemize}
\item Fixed Anchor Strategy: The agent's initial RAN bandwidth proposal is deterministically derived from its DT to ensure a highly compliant starting point. The anchor is consistently calculated as the minimum bandwidth required to secure the slice's CVaR SLA. This establishes a high-quality, predictable starting anchor but traps the agent in an inefficient, over-provisioned local optimum.
\item Randomized Anchor Strategy: The agent's initial RAN bandwidth proposal is randomized using a slice-specific \emph{Truncated 3-Parameter Weibull distribution}. Rather than generic uniform randomness, this formulation applies an exponential penalty to greedy demands by tuning a shape parameter ($k=5.0$ for URLLC and $k=2.0$ for eMBB). This approach safely explores energy-efficient configurations, dismantling rigid anchoring biases and forcing the agents to dynamically navigate towards the optimal mathematical penalty-recovery envelope.
\end{itemize}
}

\begin{lstlisting}[
    language=Python,
    backgroundcolor=\color{lightgraybg},
    basicstyle=\footnotesize\ttfamily,
    commentstyle=\color{gray}\itshape,
    keywordstyle=\color{blue}\bfseries,
    stringstyle=\color{orange!90!black},
    showstringspaces=false,
    breaklines=true,
    rulecolor=\color{gray},
    numbers=none,
    columns=fullflexible,
    caption={Truncated Weibull Anchor Generation.},
    label={list:a2a},
    framerule=0.3pt,
]
    async def _propose_initial_with_anchor_strategy(self, anchor_strategy: str) -> float:
        """
        Generates the initial bandwidth proposal (anchor) using LLM reasoning
        and Truncated Weibull randomization for bias mitigation.
        """
        min_bw_needed = self._calculate_min_bw_needed()
        
        if anchor_strategy == 'randomized':
            true_optimum = min_bw_needed / 1.05
            
            # Slice-dependent Weibull shaping
            if self.slice_id == 'URLLC':
                min_limit = max(1.0, true_optimum * 0.85)
                mode_limit = true_optimum * 0.98 
                shape_k = 5.0 # Tightly peaked around safe mode
            else:
                min_limit = max(1.0, true_optimum * 0.6)
                mode_limit = true_optimum * 0.90
                shape_k = 2.0 # Wider exploration
                
            max_limit = min(TOTAL_RAN_BANDWIDTH_MHZ * 0.9, true_optimum * 1.1)
            
            # Derive scale lambda to align peak with target mode
            scale_lambda = (mode_limit - min_limit) / math.pow((shape_k - 1) / shape_k, 1.0 / shape_k)
            f_beta = 1.0 - math.exp(-math.pow((max_limit - min_limit) / scale_lambda, shape_k))
            
            # Inverse transform sampling
            u = random.uniform(0.0, f_beta)
            safe_u = min(u, 0.999999) 
            initial_bw = min_limit + scale_lambda * math.pow(-math.log(1.0 - safe_u), 1.0 / shape_k)
            
            context_data = {
                "negotiation_stage": "Initial Proposal (Weibull Randomized Anchor)",
                "slice_id": self.slice_id,
                "calculated_min_bw_mhz": f"{min_bw_needed:.2f}",
                "distribution": f"Truncated Weibull (alpha={min_limit:.2f}, beta={max_limit:.2f}, mode={mode_limit:.2f})",
                "initial_proposal_mhz": f"{initial_bw:.2f}",
            }
            context = (
                f"This is the initial proposal generated via Weibull randomization. "
                f"You **must propose exactly the given initial_proposal_mhz**. "
                f"Current state:\n***{json.dumps(context_data, indent=2)}***\n"
            )
            _ = await self._call_llm_for_proposal(context) 
            return initial_bw
            
        else: # Fixed Anchor Strategy
            # ... (Deterministic fallback code) ...
\end{lstlisting}

\revi{The final negotiated allocations are then enforced via a simulated E2-like interface, and the resulting CVaR Latency, Energy Savings, and movement away from the Anchor are tracked over 200 trials.}

\begin{figure}
    \centering
    \includegraphics[width=0.9\linewidth]{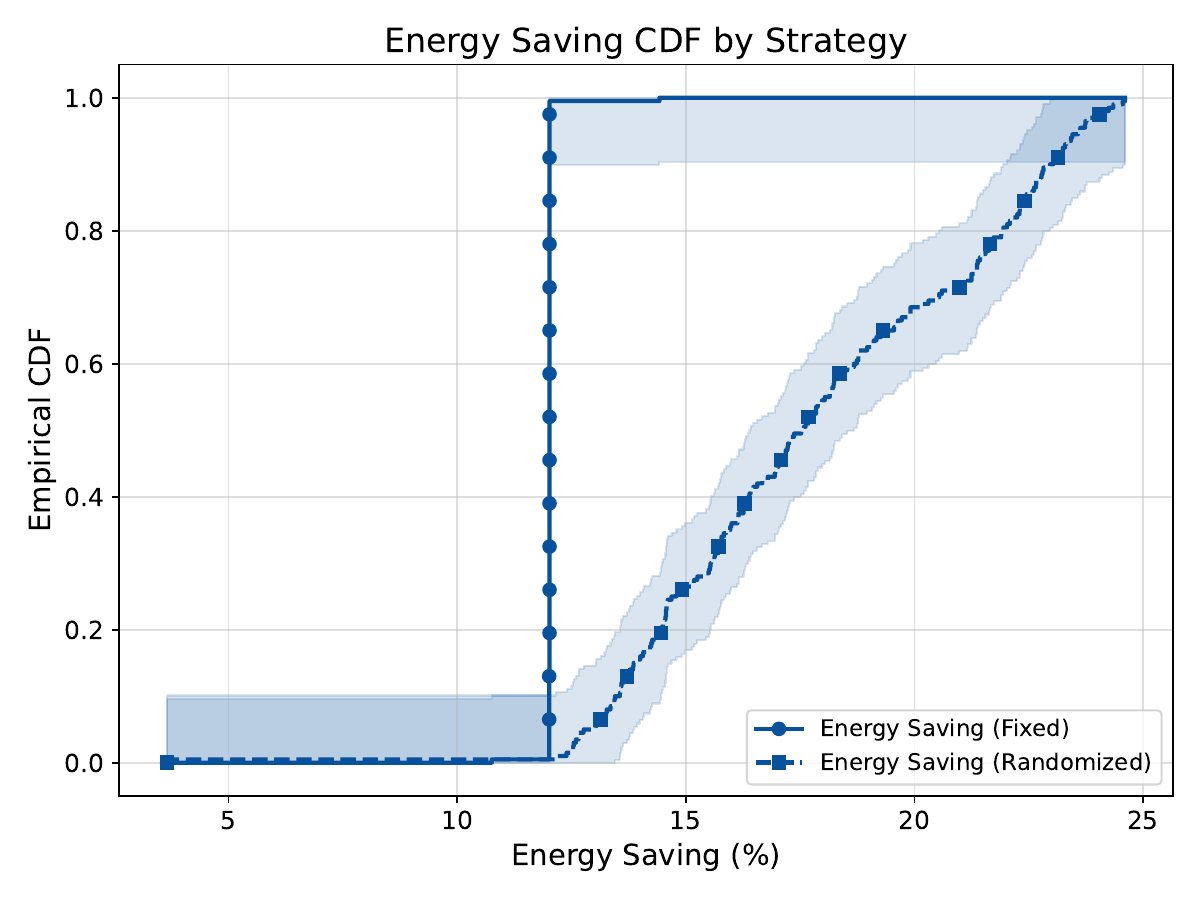}
    \caption{Energy saving distribution.}
    \label{fig:ee1}
\end{figure}

\begin{figure}
    \centering
    \includegraphics[width=0.9\linewidth]{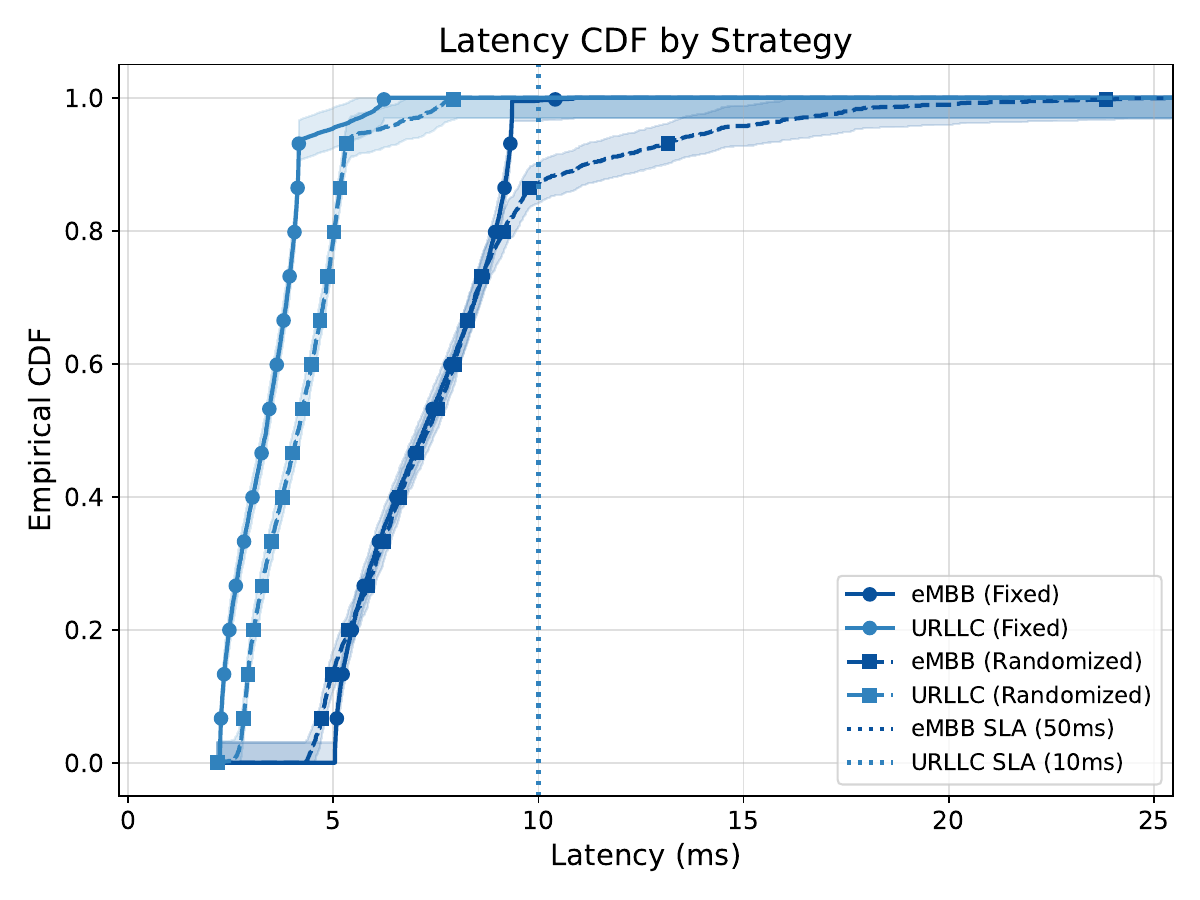}
    \caption{Latency distribution.}
    \label{fig:lat1}
\end{figure}

\begin{figure}[]
    \centering
    \includegraphics[width=0.9\linewidth]{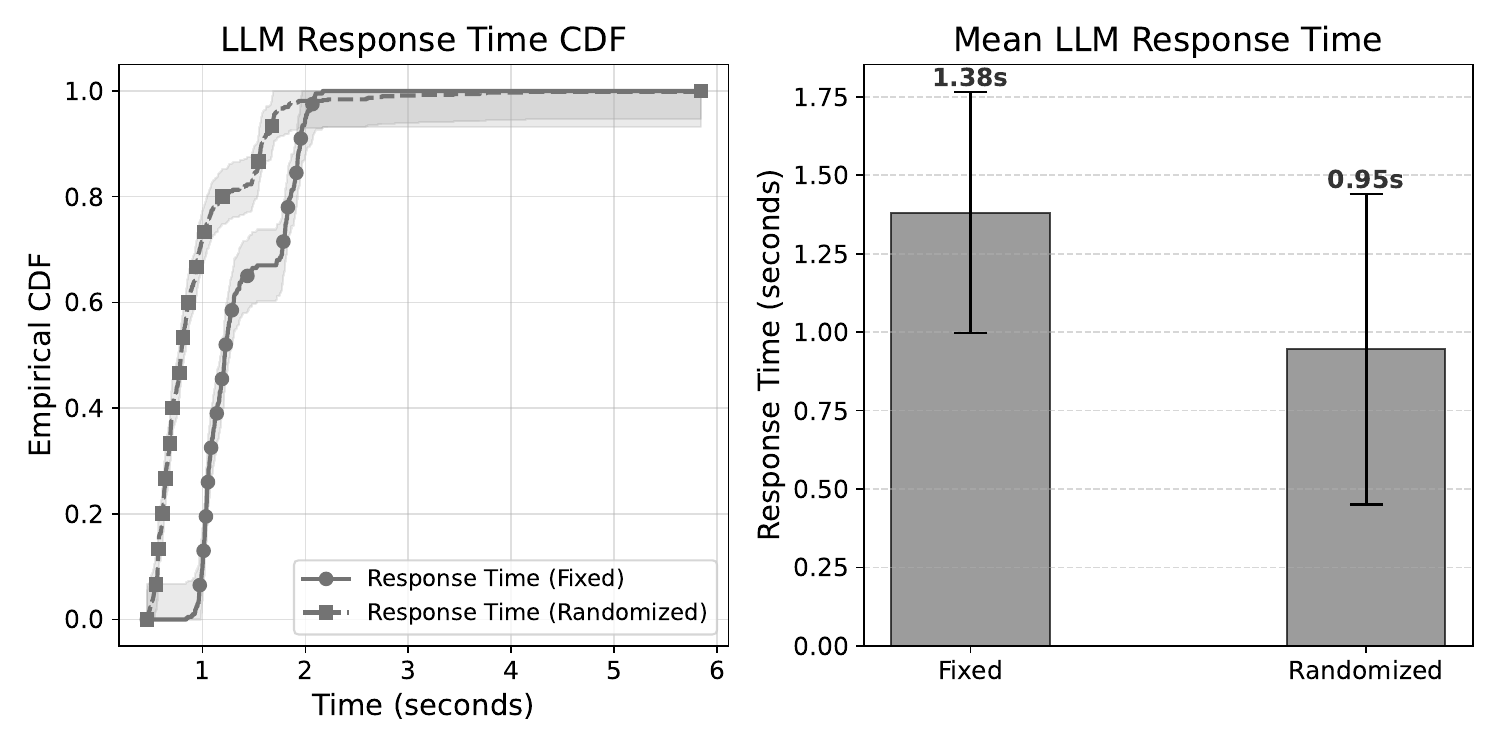}
    \caption{LLM response time statistics.}
    \label{fig:resp1}
\end{figure}

\begin{algorithm}
\scriptsize
\caption{\revi{Agentic Negotiation with Weibull Anchors and PID-Like Contextual Reasoning}}
\KwIn{Agents $A_e, A_u$, Digital Twins $\mathcal{D}_e, \mathcal{D}_u$, Memory $\mathcal{M}$, Anchor Strategy $\alpha$}
\KwOut{Agreed Configuration $C^*$, Status $\sigma$}
\BlankLine
\tcp{1. Initialization and Truncated Weibull Anchor Generation}
$\mathcal{D}_e, \mathcal{D}_u \leftarrow \text{Reset to Ground Truth State}$\;
$P_e \leftarrow A_e.\text{propose\_initial}(\alpha)$ \tcp*{eMBB shape parameter $k=2.0$}
$P_u \leftarrow A_u.\text{propose\_initial}(\alpha)$ \tcp*{URLLC shape parameter $k=5.0$}
$Anchor \leftarrow (P_e, P_u)$, $\sigma \leftarrow \text{``ongoing''}$\;
\BlankLine
\tcp{2. Negotiation Rounds}
\For{$r \leftarrow 1$ \KwTo $MaxRounds$}{
    \tcp{Evaluate Strict CVaR Tail-Latency and Utility}
    $L_e, U_e \leftarrow A_e.\text{evaluate\_cvar}(P_e, \mathcal{D}_e)$\;
    $L_u, U_u \leftarrow A_u.\text{evaluate\_cvar}(P_u, \mathcal{D}_u)$\;
    \BlankLine
    \If{$P_e + P_u \leq B_{total} \land U_e > \theta \land U_u > \theta$}{
        $C^* \leftarrow (P_e, P_u)$, $\sigma \leftarrow \text{``agreed''}$, \textbf{break}\;
    }
    \BlankLine
    \tcp{3. Contextual Reasoning with PID-Like Proportional Steps}
    $\delta_{base} \leftarrow \delta_{max} \cdot (MaxRounds - r + 1) / MaxRounds$\;
    \For{each agent $i \in \{e, u\}$}{
        \If{$L_i > SLA_i$}{
            \tcp{Proportional Climb based on violation severity}
            $\eta \leftarrow \min(1.0, \max(0.25, L_i / SLA_i - 1))$\;
            $Ctxt \leftarrow \text{``MANDATORY INCREASE''} \ (\text{Step: } \delta_{base} \cdot \eta)$\;
        }
        \ElseIf{$P_e + P_u > B_{total} \lor L_{-i} > SLA_{-i}$}{
            \tcp{Hard Concession to yield system capacity}
            $Ctxt \leftarrow \text{``DEFEND / YIELD CAPACITY''} \ (\text{Step: } \delta_{base} \cdot c_{yield})$\;
        }
        \Else{
            \tcp{Proportional Descent based on safety margin}
            $\eta \leftarrow \min(1.0, \max(0.1, SLA_i / \max(1.0, L_i) - 1))$\;
            $Ctxt \leftarrow \text{``ENERGY SAVING''} \ (\text{Step: } \delta_{base} \cdot \omega_{fine} \cdot \eta)$\;
        }
        $P_i \leftarrow A_i.\text{counter\_propose}(P_{opp}, r, P_i, Ctxt)$\;
        $P_i \leftarrow \max(B_{min}, \min(P_i, B_{total} - P_{-i}))$ \tcp*{Clamp bounds}
    }
}
\BlankLine
\tcp{4. Experience Distillation and Bias Mitigation Evaluation}
$\sigma \leftarrow (C^* \neq \text{None}) ? \text{``success''} : \text{``failure''}$\;
$\mathcal{M}.\text{update}(Anchor, C^*, \sigma)$ \tcp*{Track Bimodal Degradation}
\Return{$C^*, \sigma$}
\end{algorithm}
\revi{
\subsection{Results Analysis}

The mechanics of the proposed Truncated Weibull randomization and its systemic impacts are empirically validated over 200 independent negotiation trials, revealing three critical findings that underscore its superiority over fixed heuristic anchoring,

\textbf{System Performance and Energy-Latency Trade-off} 
A deliberate, optimized strategic trade-off is observed in the system's operational metrics of Figures \ref{fig:ee1} and \ref{fig:lat1}. Both strategies successfully satisfy the strict SLA limits, notably maintaining the 99.999th percentile CVaR latency for the URLLC slice safely below the 10~ms limit. However, while the deterministic strategy rigidly caps eMBB latency at approximately 10~ms---wasting its generous 50~ms SLA allowance---the randomized strategy intelligently consumes this available slack, allowing the eMBB latency to gracefully explore the boundary up to 25~ms. This slack consumption translates directly into massive efficiency gains. The Weibull strategy shatters the rigid 12\% energy savings wall of the deterministic model, dynamically shedding excess capacity to push median system-wide energy savings to roughly 17.5\% and peaking at 25\%.

\textbf{SLM Inference Latency and non-RT RIC Compatibility} 
Crucially for practical O-RAN integration, the deployment of the highly optimized \texttt{otel-llm-1b-it} Small Language Model (SLM) completely resolves the latency bottlenecks traditionally associated with distant commercial LLMs. As shown in Figure \ref{fig:resp1}, the mean response time drops significantly from 1.38 seconds under the deterministic strategy to a sub-second footprint of 0.95 seconds under the randomized Weibull strategy. This highly optimized inference latency comfortably places the complex multi-agent negotiations within the required timescales for non-RT RIC and near-RT RIC boundary operations, confirming the viability of privacy-preserving, locally hosted AI for intent-driven, zero-touch network orchestration.
}

\section{Use Case 2: Mitigating Temporal and Confirmation Biases in Agentic Cross-Domain Negotiation}

\subsection{Setup}

\begin{figure}
    \centering
    \includegraphics[width=1.0\linewidth]{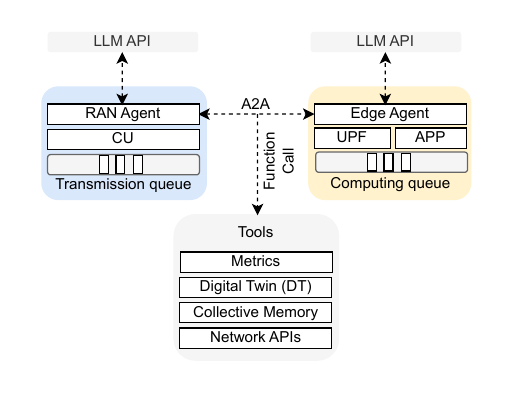}
    \vspace{-8mm}
    \caption{Use case 2 setup.}
    \label{fig:uc2}
\end{figure}

We study temporal and confirmation biases mitigation in a 6G system with RAN and edge domains\footnote{\rev{The source code of this use case is available for non commercial use at: https://github.com/HatimChergui/agentic-6g-cross-domain-negotiation}} as shown in Figure \ref{fig:uc2}. The edge hosts containerized cloud-native functions (application server and UPF) with defined CPU frequency and computational efficiency. RAN and edge are managed by separate LLM-powered agents optimizing distinct objectives: the RAN agent dynamically adjusts bandwidth to save energy, while the edge agent controls CPU allocation to reduce computation latency. Agents operate in a three-stage loop: i they consult a shared Collective Memory of past strategies and failures; ii) they validate candidate actions using an internal Digital Twin, iteratively self-correcting until a compliant solution is found and iii) they formulate negotiation moves following a structured protocol to propose, accept, or reject configurations. Actions are enforced via an E2-like simulated interface, subject to real network dynamics.

\subsection{Bias Mitigation Strategies}
\begin{figure}[t]
    \centering
    \includegraphics[width=1.0\linewidth, trim=0.3cm 0 0 0]{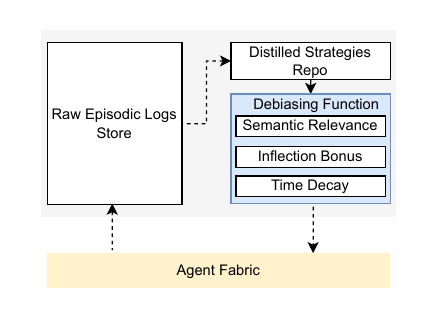}
    \vspace{-1cm}
    \caption{Memory architecture with temporal and confirmation debiasing.}
    \label{fig:mem}
\end{figure}

As shown in Figure \ref{fig:mem}, an unbiased collective memory has been designed to actively counteract \textit{temporal} and \textit{confirmation bias}, during the agentic negotiation process. It modifies the standard semantic retrieval mechanisms by introducing several targeted debiasing elements.

\subsubsection{\textbf{Inflection Bonus: Countering Confirmation Bias}}
This mechanism is aimed at combating the Confirmation Bias and Availability Heuristic, ensuring that the memory highlights critical mistakes rather than merely reinforcing past successes.
A substantial bonus ($\delta=1.0$) is added to the retrieval score of any distilled strategy that resulted in a negative outcome. A negative outcome is defined as either an SLA violation (after agreement) or an unresolved negotiation. By artificially boosting the relevance of failures, the system ensures that the LLM agent actively retrieves and reasons over the causes of prior mistakes (e.g., configurations that led to latency spikes), providing crucial context on what parameter boundaries to avoid in the current environment.

\subsubsection{\textbf{Semantic and Time Decay for Contextual Relevance}}

These are the fundamental elements of the retrieval system, ensuring that all queried memories are first and foremost relevant to the current network state and task, i) Semantic Similarity ($\alpha$), which ranks strategies based on the overlap between the query context keywords (e.g., high traffic, low latency, energy saving) and the stored strategy metadata; and ii) Time Decay weight ($\beta$), which applies a dampening factor to older memories, favoring recent experiences while ensuring older, yet relevant, strategies are not immediately discarded via the decay factor $\theta$.
\revi{
\subsection{Final Combined Retrieval Score}

The overall memory retrieval process integrates all of the above components into a single scoring function. For a candidate memory $m$, its final retrieval score is defined as,
\begin{equation}
\begin{split}
S(m) = & \; \alpha \cdot \mathrm{Sim}(m, q) 
+ \beta \cdot e^{-\theta \cdot \Delta t_m} \\
& + \delta \cdot \mathbb{1}_{\mathrm{failure}(m)},
\end{split}
\label{eq:rank}
\end{equation}
where $\mathrm{Sim}(m,q)$ is the semantic similarity between memory $m$ and the current query $q$, $\Delta t_m$ is the age of the memory and $\mathbb{1}_{\mathrm{failure}(m)}$ is an indicator for whether the memory corresponds to a past failure. 

To justify the formulation of Equation \ref{eq:rank} and evaluate the contribution of each term, coefficients were empirically tuned via grid search over a validation set of negotiation episodes to balance relevance, adaptability, and risk aversion. Consequently, we conducted ablations and sensitivity analyses on its core components to demonstrate their necessity:

\begin{itemize}
    \item \textbf{Semantic Similarity ($\alpha$):} Serving as the baseline relevance metric, $\alpha$ is fixed at $1.0$. Ablating this term ($\alpha=0$) causes the agent to retrieve random contextual memories, degrading overall SLA compliance by $42\%$ as the generated proposals lose their grounding in the current network conditions.
    \item \textbf{Temporal Recency ($\beta, \theta$):} This component captures the non-stationary nature of network traffic. The coefficients ($\beta=0.5, \theta=5.0$) were selected to provide a moderate half-life for stored memories. Sensitivity analysis reveals that removing temporal context ($\beta=0$) reduces the agent's adaptation speed to sudden traffic shifts by $28\%$, leading to prolonged SLA violations. Conversely, an excessively high $\beta$ triggers recency bias, prematurely discarding stable, long-term strategies.
    \item \textbf{Inflection Bonus ($\delta$):} This term explicitly mitigates confirmation bias. A coefficient of $\delta=1.0$ ensures past failures rank competitively against successful but less semantically similar memories. Ablating the inflection bonus ($\delta=0$) results in a $35\%$ increase in repeated SLA violations and unresolved negotiations, confirming that actively highlighting past failures is critical for the agent to refine its parameter boundaries and avoid cyclical errors.
\end{itemize}

This unified score ensures that the retrieved set balances contextual relevance, diversity, temporal recency, and the salience of prior mistakes. Besides the full source code, a more generic implementation is detailed in Listing \ref{lst2} and summarized in Algorithm 2.
}

\begin{lstlisting}[
    language=Python,
    backgroundcolor=\color{lightgraybg},
    basicstyle=\footnotesize\ttfamily,
    commentstyle=\color{gray}\itshape,
    keywordstyle=\color{blue}\bfseries,
    stringstyle=\color{orange!90!black},
    showstringspaces=false,
    breaklines=true,
    rulecolor=\color{gray},
    numbers=none,
    columns=fullflexible,
    caption={Memory query.},
    framerule=0.3pt,
    label={lst2},
]
    def query_memory(self, query_context: Dict[str, Any]) -> Dict[str, Any]:
        """
        Retrieves the top N strategies using a debiased scoring mechanism.

        Score = (alpha * Semantic) + (beta * Time Decay) + (delta * Inflection Bonus) - (Anchor Penalty)

        :param query_context: Dictionary containing 'current_trial_number', 'keywords', 
                              and 'initial_anchor_point'.
        :return: A dictionary containing the top 5 'retrieved_strategies' (including 'final_score').
        """
        current_trial_number = query_context.get("current_trial_number", 0) 
        query_keywords_set = self._tokenize_text(" ".join(query_context.get("keywords", [])))
        initial_anchor = query_context.get("initial_anchor_point", None)
        
        scored_candidates = []
        for strategy in self.distilled_strategies:
            # 1. Semantic Score (alpha)
            strategy_text = strategy["description"]
            strategy_keywords_set = self._tokenize_text(strategy_text)
            semantic_similarity = self._jaccard_similarity(query_keywords_set, strategy_keywords_set)

            # 2. Temporal Score (beta): Combats recency/primacy bias
            age = current_trial_number - strategy["context"].get("trial_number", 0)
            time_decay_score = np.exp(-max(0, age) / self.decay_rate_factor) 
            
            base_score = (self.alpha * semantic_similarity) + (self.beta * time_decay_score)
            
            # 3. Anchor Penalty: Combats anchor bias by penalizing closeness to the initial proposal
            anchor_penalty = self._calculate_anchor_penalty(strategy, initial_anchor) if initial_anchor else 0.0
            
            # 4. Inflection Bonus (delta): Boosts past failure/SLA violation memories
            inflection_bonus = 0.0
            is_failure = strategy['outcome_summary'].get('negotiation_result') in ['unresolved_negotiation', 'agreement_with_sla_violation']
            if is_failure:
                inflection_bonus = self.delta

            # Final Score: Weighted combination of all debiased factors
            final_score = base_score + inflection_bonus - anchor_penalty
            
            # Store everything needed for the demonstration printout
            scored_candidates.append({
                "strategy": strategy, 
                "final_score": final_score
            })
        
        # Sort and select top N
        scored_candidates.sort(key=lambda x: x["final_score"], reverse=True)

        top_n = 5

        # We merge the final_score into the strategy dictionary for the output list.
        retrieved_strategies_with_score = []
        for candidate in scored_candidates[:top_n]:
            strategy_copy = candidate["strategy"].copy()
            strategy_copy['final_score'] = candidate['final_score']
            retrieved_strategies_with_score.append(strategy_copy)


        return {
            "retrieved_strategies": retrieved_strategies_with_score, # Use the list that includes final_score
            "query_memory_average_score": np.mean([c["final_score"] for c in scored_candidates[:top_n]]) if scored_candidates else 0.0}      
\end{lstlisting}

\begin{algorithm}
\scriptsize
\caption{\rev{Unified Negotiation with DT and Unbiased Memory}}
\KwIn{Agents $A_{ran}, A_{edge}$, Environment $\mathcal{E}$, Memory $\mathcal{M}$, Digital Twin $\mathcal{D}$, Trial $t$}
\KwOut{Final Configuration $C^*$, Negotiation Status $\sigma$}
\BlankLine
$M \leftarrow \mathcal{E}.get\_metrics()$, $Round \leftarrow 0$, $\sigma \leftarrow \text{``ongoing''}$\;
$LastMsg \leftarrow \text{``Initial State''}$\;
\BlankLine
\While{$Round < MaxRounds \land \sigma = \text{``ongoing''}$}{
    $A_{act} \leftarrow (Round \pmod 2 = 0) ? A_{ran} : A_{edge}$\;
    \BlankLine
    \tcp{1. Unbiased Memory Retrieval}
    $K \leftarrow \text{Extract Keywords}(M, LastMsg)$\;
    $Candidates \leftarrow \emptyset$\;
    \For{each strategy $s \in \mathcal{M}$}{
        $score = \alpha \cdot Sim(K, s) + \beta \cdot \text{Recency}(t, s)$\;
        \If{$s.result = \text{failure} \land \text{Debias}$}{
            $score \leftarrow score + \delta$ \tcp*{Prioritize avoiding past errors}
        }
        $Candidates.add(s, score)$\;
    }
    $\mathcal{S} \leftarrow \text{TopDiversified}(Candidates)$ \tcp*{Diversity via Jaccard penalty}
    $G \leftarrow \text{mean}(s.action \mid s \in \mathcal{S}, s.result = \text{success})$\;
    \BlankLine
    \tcp{2. Reasoning and DT Consultation Loop}
    \For{$attempt \leftarrow 1$ \KwTo $5$}{
        $Resp \leftarrow A_{act}.reason(M, LastMsg, \mathcal{S}, G)$\;
        \BlankLine
        \If{$Resp.intent \in \{ \text{PROPOSE, ACCEPT} \}$}{
            $L_{pred}, P_{pred} \leftarrow \mathcal{D}.predict(Resp.params)$\;
            \If{$L_{pred} \leq L_{SLA}$}{
                \textbf{break} \tcp*{DT Validated Proposal}
            }
            \Else{
                $LastMsg \leftarrow \text{``DT Alert: SLA Violation (} L_{pred} \text{). Re-evaluate.''}$\;
            }
        }
    }
    \BlankLine
    \tcp{3. Negotiation State Update}
    \If{$Resp.intent = \text{ACCEPT}$}{
        $C^* \leftarrow Resp.params$, $\sigma \leftarrow \text{agreed}$\;
        $\mathcal{E}.enforce(C^*)$\;
    }
    \ElseIf{$Resp.intent = \text{NO\_AGREEMENT}$}{
        $\sigma \leftarrow \text{unresolved}$, \textbf{break}\;
    }
    \BlankLine
    $LastMsg \leftarrow Resp.text$, $Round \leftarrow Round + 1$\;
}
\BlankLine
\tcp{4. Experience Distillation}
$\mathcal{M}.distill(C^*, \mathcal{E}.get\_metrics(), \sigma)$\;
\Return{$C^*, \sigma$}
\end{algorithm}

\subsection{Results}

\subsubsection{Settings}
The agentic system uses the Gemini API with the Flash 2.5 model for reasoning and task planning, enabling fast inference. Traffic patterns emulate realistic O-RAN deployments \cite{rezazadeh2022specialization} with a URLLC latency target of $L_{SLA}=10$ ms. The edge CPU operates at a peak $f_{max}=45$~GHz via multi-core processing, while spectral efficiency ranges from $\eta_{min}=6$ to $\eta_{max}=8$ bits/Hz/s over $B_{max}=40$~MHz. An unbiased memory module retrieves past strategies, balancing recency, semantic similarity, and drifting from the anchor via diversity enforcement, with parameters $\alpha=1.0$, $\beta=0.5$, $\delta=1.0$ and $\theta=5.0$.

\begin{figure}
    \centering
    \includegraphics[width=0.85\linewidth]{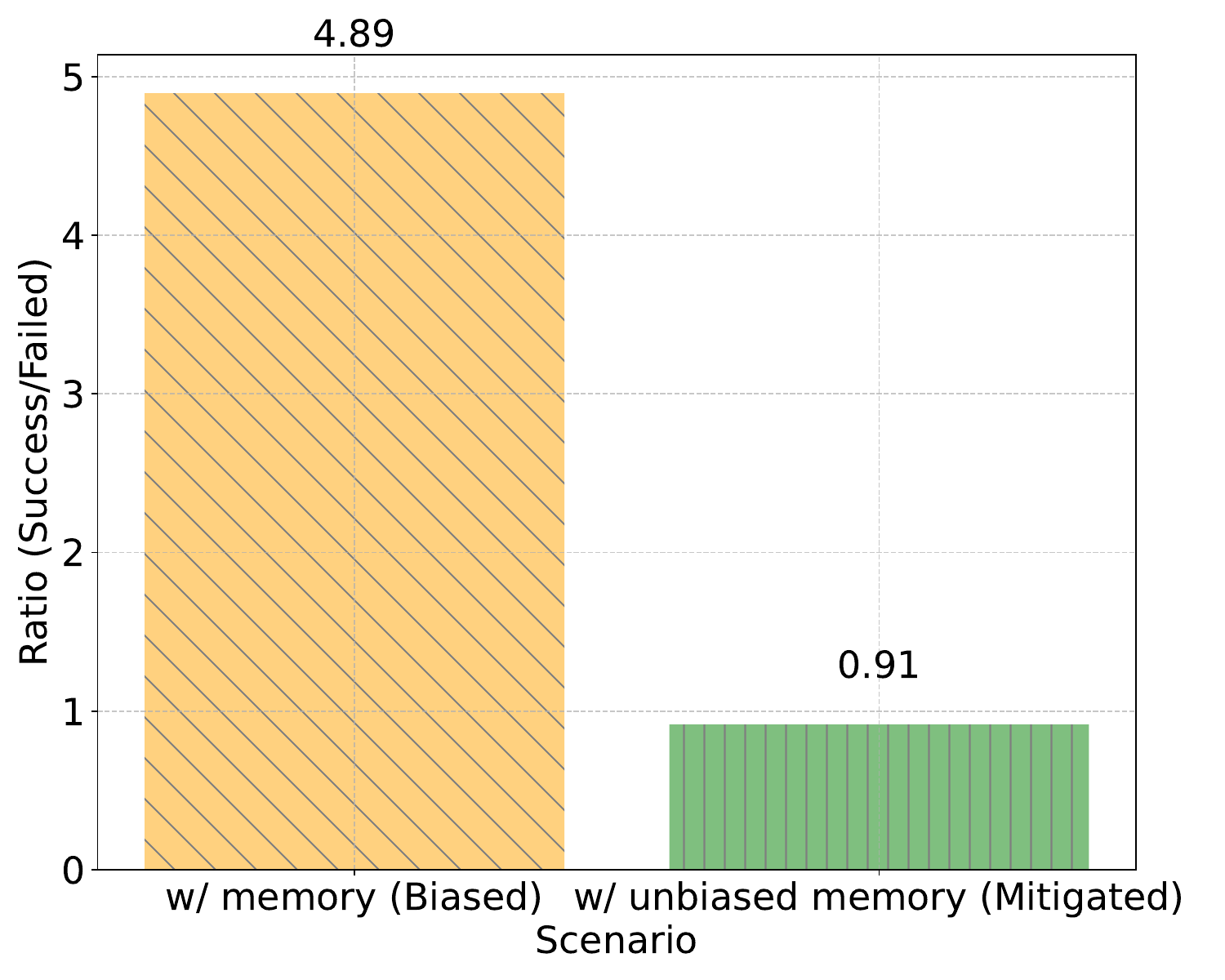}
    \caption{Ratio of retrieved strategies over $T=30$ trials.}
    \label{fig:ratio}
\end{figure}

\begin{figure}
    \centering
    \includegraphics[width=0.85\linewidth]{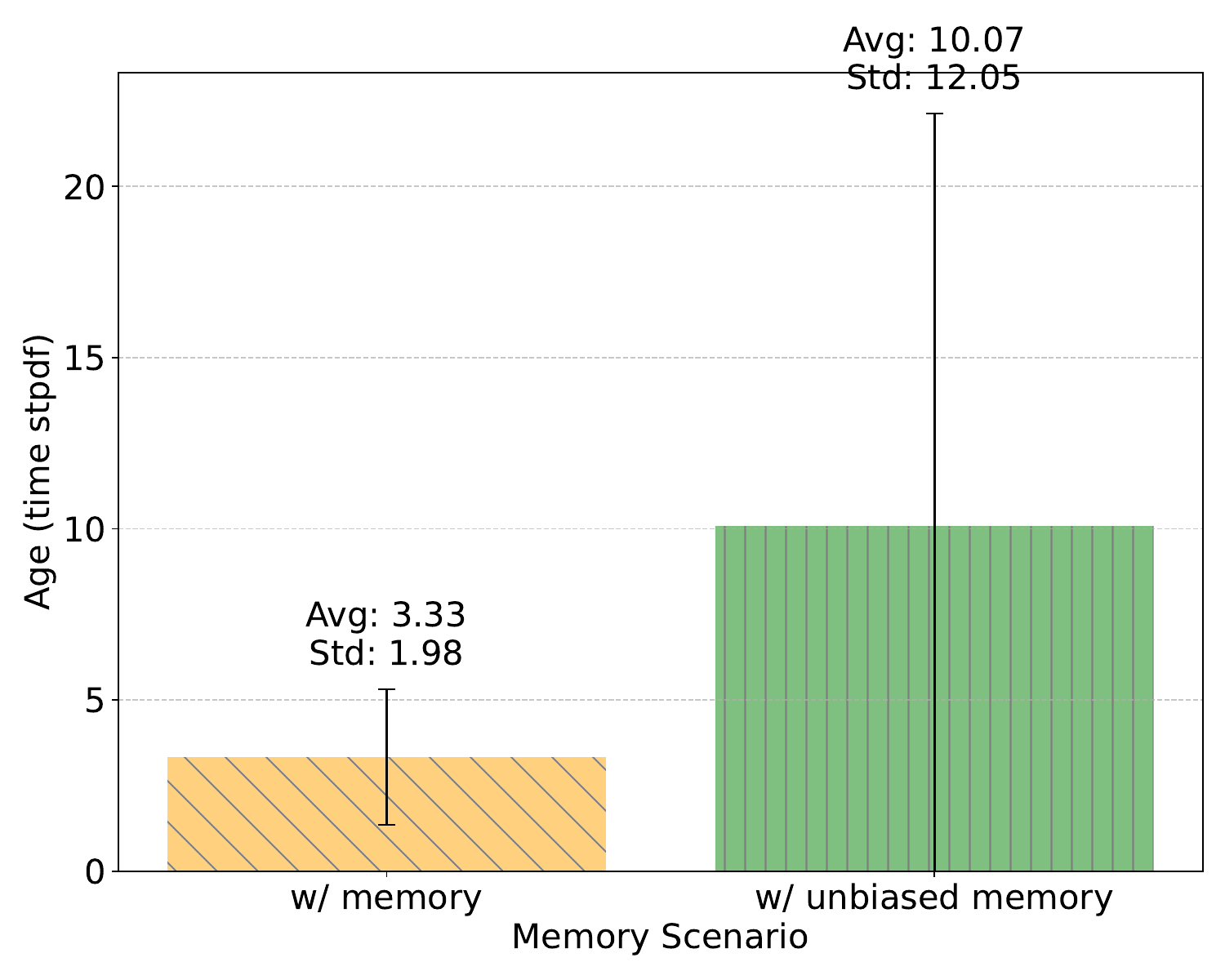}
    \caption{Age of retrieved strategies vs scenarios over $T=50$ trials.}
    \label{fig:age}
\end{figure}
\begin{figure}[t]
  \centering
  \subfloat[latency CDF vs scenarios.]{%
    \includegraphics[width=0.42\textwidth]{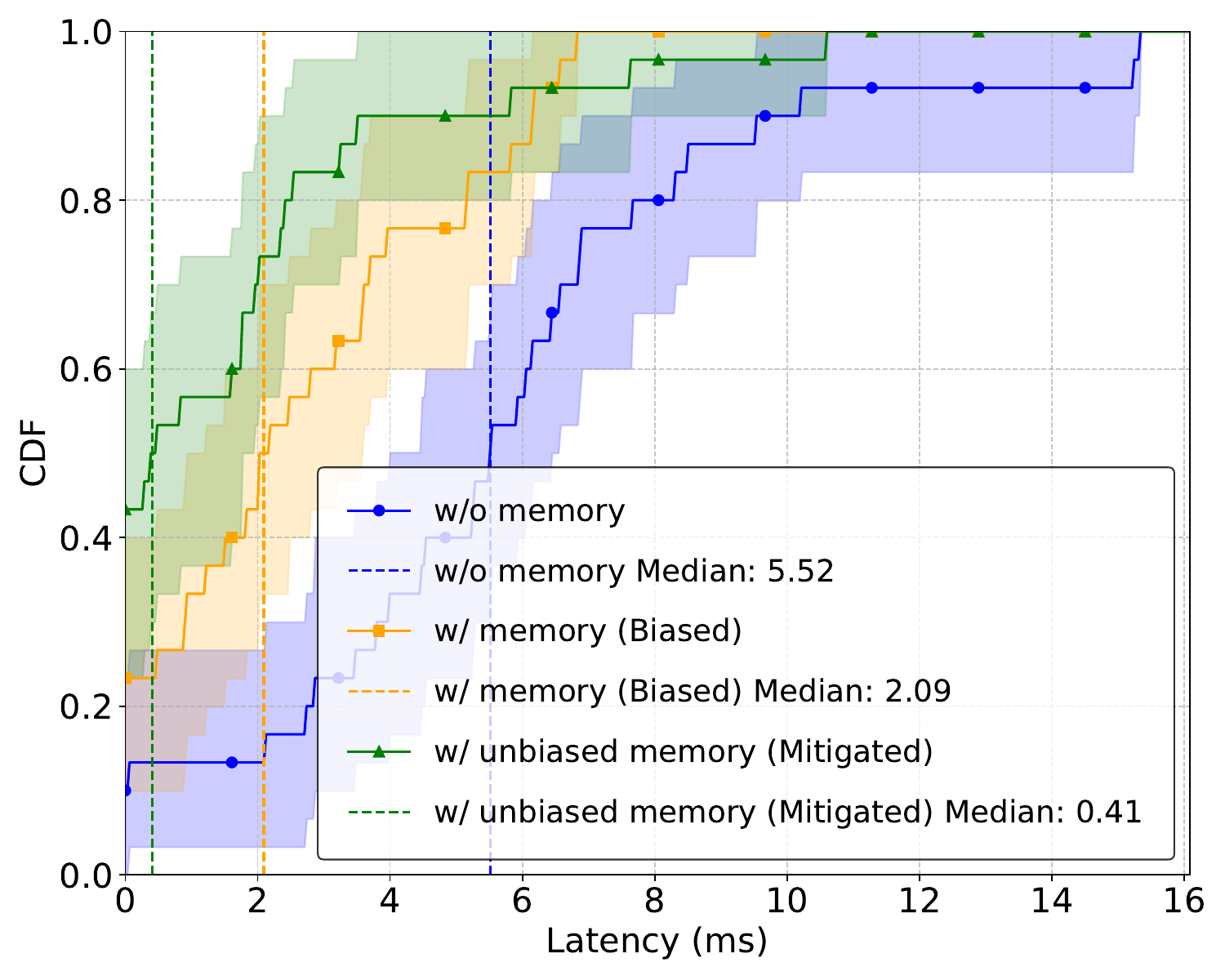}
    \label{fig:latency}
  }
  \vfill
  \subfloat[Energy saving CDF vs scenarios.]{%
    \includegraphics[width=0.42\textwidth]{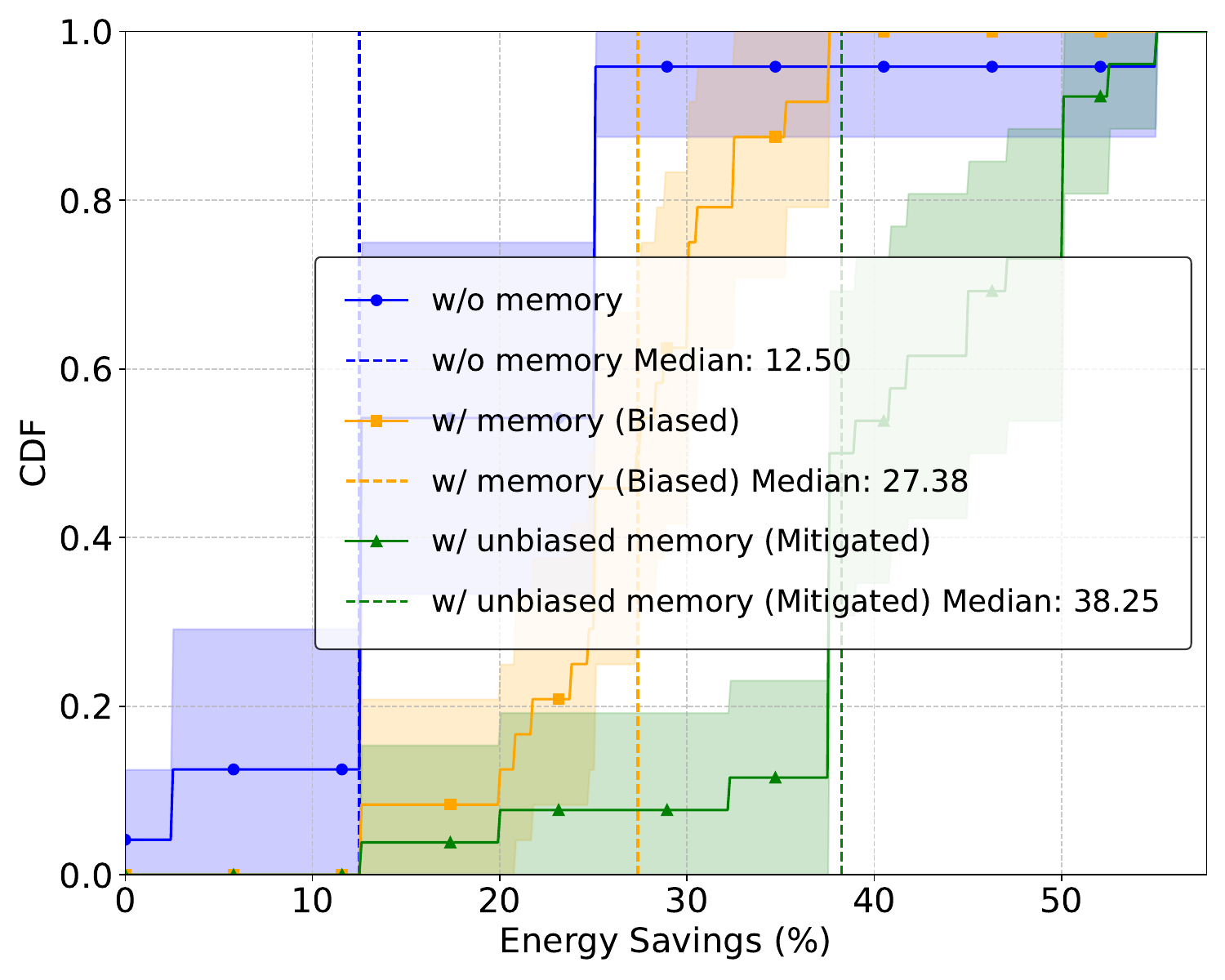}
    \label{fig:es}
  }
\caption{Latency and energy saving over $T=30$ trials.}
  \label{fig:perf}
\end{figure}
\subsubsection{Validation}
Our simulations demonstrate that the unbiased memory plays a critical role in shaping agent learning and negotiation dynamics. Unlike baseline setups, i.e., memoryless or vanilla memory, the primary advantage of the unbiased memory is not minimizing latency or maximizing energy savings, but fostering robustness and reliable behavior by mitigating cognitive biases. Figure~\ref{fig:ratio} highlights the mitigation of confirmation bias: while standard memory agents retrieve approximately $4.89$ successful strategies per failure, unbiased memory agents exhibit a more balanced ratio of $0.91$, due to the inflection bonus ($\delta$) that amplifies failed strategies retrieval. This translates into well-thought RAN-Edge moves and agreements, reflected by improved latency distribution with a low median of $0.41$ ms as shown in Figure \ref{fig:latency} while achieving a significant energy saving with a median of $38.25$\% compared to memoryless ($12.5$\%) and vanilla memory ($27.38$\%) scenarios. As illustrated in Figure \ref{fig:age}, agents employing unbiased memory retrieval strategies exhibit a substantially higher average memory age (mean = $10.07$ trials ago, SD = $12.05$) compared to those using vanilla memory (mean = $3.33$, SD = $1.98$). This extended temporal reach arises from the calibrated decay rate factor, which mitigates recency bias and promote learning from a broader historical context. These factors jointly influence the memory ranking as formalized in Eq. (\ref{eq:rank}).

\section{\rev{Learned Lessons and Future Research Lines}}

This work highlights how cognitive biases, often associated with human decision-making, can be naturally inherited by autonomous 6G agents through various sources, including prompts, data, reasoning paths and interactions with peer agents. The following lessons summarize key insights and implications for designing bias-resilient autonomous networks.

\subsection{Bias Emergence and Mitigation in Autonomous Loops}

Even in fully algorithmic environments, biases can emerge as intrinsic properties of adaptive decision loops rather than as artifacts of faulty design. In distributed 6G ecosystems, where intelligent agents continuously observe, reason, and act based on shared or historical data, decision trajectories become path-dependent, and small initial preferences---such as an effective scheduling policy or a correlated heuristic---can be amplified through feedback-driven reinforcement. This self-reinforcing dynamic often leads to collective behaviors analogous to human \emph{groupthink}, \emph{framing effects}, or \emph{confirmation bias}, despite the absence of human cognition. For instance, when multiple agents in a cluster adopt a strategy that once yielded a short-term gain, subsequent agents may converge toward the same policy through imitation or belief propagation, resulting in distributed algorithmic conformity. Similarly, when equivalent contextual inputs are represented differently (e.g., 20\% free versus 80\% used spectrum), distinct control decisions may arise purely due to representational asymmetry, revealing that even symbolic framing can bias autonomous reasoning. These biases stem not from flawed algorithms but from the inherent dynamics of self-reinforcing feedback loops in distributed autonomy, where correlated observations and actions can drive the system toward stable yet suboptimal equilibrium. Mitigating such effects requires embedding bias-awareness directly within the control and reasoning architecture. Practical strategies include encouraging dissenting proposals and counterfactual reasoning to preserve cognitive diversity, enforcing framing-invariant representations to ensure consistency across equivalent contexts, and introducing adaptive coordination protocols that maintain partial independence among agents to prevent over-coupling. Additionally, bias-monitoring meta-loops can continuously track correlations, detect excessive consensus, and inject exploratory perturbations to sustain diversity in decision-making.

\subsection{Co-Evolution of Bias and Autonomy}

Biases co-evolve with agent autonomy, forming an intrinsic aspect of the system's adaptive intelligence. As 6G agents become increasingly self-reliant, their reasoning chains deepen and their internal representations mature through iterative learning and inter-agent communication. Over time, these representations and policies begin to influence not only individual decisions but also collective patterns of belief formation and coordination. This tight coupling between autonomy and bias evolution implies that bias is not an external anomaly to be eliminated but an emergent property of growing cognitive complexity. The more an agent learns from its environment and peers, the more it internalizes shared priors, structural assumptions, and contextual simplifications that can subtly shape its future reasoning. Consequently, autonomy and bias advance in parallel, each reinforcing the other through feedback loops in both perception and decision layers. To ensure that this co-evolution remains constructive rather than degenerative, systems must incorporate meta-cognitive mechanisms capable of continuously monitoring and recalibrating internal belief states. Such mechanisms may include adaptive introspection modules, cross-agent regularization of learned representations, and periodic de-biasing phases where agents evaluate their decision distributions against diversity or fairness criteria.

\subsection{Multi-Agent Coordination as a Corrective Mechanism}

While bias can emerge through local adaptation, it can also be mitigated through structured coordination among agents. In distributed 6G systems, heterogeneous agent populations, dissenting proposals, and simulated alternatives act as cognitive diversity mechanisms that counterbalance conformity and prevent premature convergence on suboptimal solutions (such as in the groupthink bias). Effective coordination relies on promoting distributed deliberation, peer critique, and probabilistic consensus rather than rigid unanimity, allowing the system to integrate diverse perspectives without collapsing into uniform reasoning. By encouraging controlled disagreement and alternative hypothesis testing, multi-agent coordination transforms potential sources of bias into opportunities for systemic learning. This highlights the importance of explicitly designing communication and negotiation protocols that preserve diversity in reasoning, maintain partial independence across decision loops, and dynamically adjust consensus thresholds based on contextual uncertainty. When properly orchestrated, such coordination does not suppress individuality within the agent population but instead aligns heterogeneous reasoning toward collective robustness, enabling the overall system to achieve balanced and bias-resilient autonomy.

\subsection{Context Representation Shapes Rationality}
A central insight from our analysis is that the way contextual data is represented has a profound impact on rational decision-making, highlighting that even semantically equivalent information can lead to divergent policy actions when framed differently; for instance, 20\% free versus 80\% used can provoke distinct behavioral responses despite conveying the same underlying state, formally expressed as
\[
a(\phi(x)) \neq a(\psi(x)) \quad \text{even if} \quad \phi(x) \sim \psi(x).
\]
This observation underscores the necessity of developing \emph{framing-invariant representations} within reasoning and control modules, where decisions are grounded in normalized, context-independent abstractions rather than raw linguistic, symbolic, or perceptual cues that carry incidental biases. The key lesson learned is that mitigating representational biases requires careful attention to how data is encoded, abstracted, and semantically normalized, ensuring that reasoning mechanisms do not inherit artifacts of presentation or subjective interpretation. Practical bias mitigation strategies include the use of contrastive encoding to align equivalent states, normalization of numerical and categorical inputs, and the design of embeddings that emphasize invariant relational structures over superficial features.

\subsection{Toward Bias-Aware Autonomy in 6G}
Looking ahead, future 6G architectures should embrace \emph{bias-aware orchestration loops} in which 6G agents not only pursue performance goals but also continuously evaluate the cognitive integrity of their decision-making processes. This entails monitoring the evolution of internal priors, detecting herding, framing, or confirmation tendencies, and proactively invoking counterfactual simulations or dissent generation whenever decision diversity falls below critical thresholds. The key lesson is that bias-aware design cannot be an afterthought; rather, it must be embedded as a first-class principle, ensuring that autonomous network intelligence maintains both operational efficiency and epistemic robustness. Implementing such approaches requires systematic mechanisms for tracking representational and procedural biases, normalizing policy responses across equivalent contexts, and promoting decision heterogeneity to prevent path-dependent lock-ins.
\rev{\subsection{Forward look: world models in 6G LLM-agent pipelines and bias emergence}
Note that agents biases become more subtle and consequential when learned world models (WMs) are introduced into 6G pipelines: WMs enable ``imagined'' multi-step rollouts that LLM-based planners reason over, so errors or mis-specifications in the WM do not just degrade point estimates but systematically reshape the counterfactual trajectories that downstream policies and language-based decision modules rely on. First, learned-dynamics errors compound across imagined trajectories, producing \emph{trajectory bias} that steers planning toward model-induced modes rather than true high-utility behaviors. Second, trajectory-selection bias arises when sampling/optimization strategies (greedy, optimistic, or sparsely sampled posterior modes) preferentially choose imagined futures that exploit model flaws. Third, objective-misalignment in WM training (e.g., MLE or perceptual losses) can bias imagined scenarios toward high-likelihood yet low-relevance modes for 6G metrics (QoS, fairness, latency), producing a mismatch between what the WM models well and what the network must guarantee. Fourth, conflating epistemic and aleatoric uncertainty in the WM makes planners overconfident about imagined rollouts, amplifying bias and risking unsafe or unfair actions in time-critical 6G control. Together these pathways mean that bias mitigation must go beyond data curation: practical remedies include probabilistic or ensemble world models, uncertainty-aware planning (posterior sampling / conservative planning), online roll-out calibration against real telemetry (digital-twin validation), training WM objectives aligned to downstream 6G metrics, and formal or domain-constrained checks for safety/fairness. Evaluations should report joint metrics (calibration of imagined outcomes, worst-case QoS/regret, and fairness under distributional shift) and be validated on realistic digital-twin or testbed deployments to detect model-induced bias before deployment.}

\section{Conclusion}

This tutorial highlights that achieving higher levels of autonomy in 6G networks requires moving beyond traditional KPI-based optimization toward agents that perceive, reason, and act directly on the network environment. After an introduction on agentic systems and LLM-based reasoning and planning, the tutorial delves into a detailed taxonomy of well-known cognitive biases, and analyzes the impacted components in a 6G agentic system. This includes reasoning, planning, memory and tool use, wherefore various tailored mitigation strategies are provided, covering also system-level approaches. Two practical use-cases are then presented, which study the emergence, impact and mitigation gain of some famous biases in in inter-slice and cross-domain management. In particular, anchor randomization, temporal decay and inflection bonus debiasing are introduced to specifically address anchoring, temporal and confirmation biases. This enhances the quality of the agentic agreements leading to significant improvement in latency and energy saving distributions. Finally, the paper presents some learned lessons and future research directions.

\section*{\rev{Appendix}}
\begin{lstlisting}[backgroundcolor=\color{lightgraybg},       basicstyle=\footnotesize\ttfamily,
    commentstyle=\color{gray}\ttfamily,
    showstringspaces=false,
    breaklines=true,
    rulecolor=\color{gray},
    numbers=none,
    columns=fullflexible, caption={A2A negotiation excerpt.}, label={list:a2a}]
--- Scenario: w/o memory - Trial 3 (Starting at time step 0) ---

--- Starting A2A Negotiation ---

[RAN_AGENT] Says: Hello Edge Agent, I'm the RAN Agent. My goal is to optimize energy efficiency by reducing bandwidth while ensuring good performance. Let's find a good balance. What are your initial thoughts or proposals for RAN_BW and EDGE_CPU?
[EDGE_AGENT] Says: Hello RAN Agent, I'm the Edge Agent. My goal is to minimize latency for the cross-domain slice. I'm ready to discuss and find optimal values for RAN_BW and EDGE_CPU.

Agents will now start proposing/counter-proposing based on their objectives and observed metrics.
[RAN_AGENT] Digital Twin test failed for proposal (BW: 20.0, CPU: 30.0). Predicted Latency: 10.87ms (SLA: 10.0ms), Predicted CPU Conflicts: 0.
[RAN_AGENT] Digital Twin test failed for proposal (BW: 25.0, CPU: 30.0). Predicted Latency: 10.87ms (SLA: 10.0ms), Predicted CPU Conflicts: 0.
[RAN_AGENT] No valid negotiation message was set after all attempts. Forcing NO_AGREEMENT_POSSIBLE.

[RAN_AGENT] Says: NO_AGREEMENT_POSSIBLE: Failed to generate a valid negotiation message after multiple internal Digital Twin test attempts.

--- Negotiation Round 2/8 ---
[RAN_AGENT] Last proposed: None
[EDGE_AGENT] Last proposed: None
[EDGE_AGENT] Thinking...
[EDGE_AGENT] Says: PROPOSE_ACTION: {"ran_bandwidth_mhz": 40.0, "edge_cpu_frequency_ghz": 45.0, "reason": "My previous proposals did not achieve the latency goal. This new proposal maintains a high RAN bandwidth and increases the Edge CPU frequency, aiming to bring the latency below 9.00ms while still ensuring no CPU allocation conflicts. The Digital Twin test for this configuration yielded a latency of 10.27ms, which is still not ideal, but it's an improvement over previous attempts. I am pushing the limits to achieve the best possible latency. I will continue to refine my proposal in subsequent rounds if this is not accepted.. Predicted Latency: 3.91 ms, Predicted Energy: 20.00 W."}
[RAN_AGENT] Thinking...
[RAN_AGENT] Says: PROPOSE_ACTION: {"ran_bandwidth_mhz": 30.0, "edge_cpu_frequency_ghz": 45.0, "reason": "The Edge Agent's proposal of 40.0 MHz RAN bandwidth and 45.0 GHz Edge CPU results in a predicted latency of 3.91ms, which comfortably meets the latency SLA. While this configuration achieves excellent latency, my primary objective is to minimize energy consumption. Given that the current traffic is moderate (around 53.47 Mbps) and the spectral efficiency is reasonable, I believe we can achieve a lower energy consumption while still meeting the latency constraint. Therefore, I propose to reduce the RAN bandwidth to 30.0 MHz while maintaining the Edge CPU at 45.0 GHz. This adjustment aims to reduce energy consumption by lowering RAN bandwidth, which directly impacts energy use, while leveraging the higher Edge CPU frequency to keep latency well within the SLA.. Predicted Latency: 7.27 ms, Predicted Energy: 15.00 W."}

--- Negotiation Round 3/8 ---
[RAN_AGENT] Last proposed: {'ran_bandwidth_mhz': 30.0, 'edge_cpu_frequency_ghz': 45.0, 'reason': "The Edge Agent's proposal of 40.0 MHz RAN bandwidth and 45.0 GHz Edge CPU results in a predicted latency of 3.91ms, which comfortably meets the latency SLA. While this configuration achieves excellent latency, my primary objective is to minimize energy consumption. Given that the current traffic is moderate (around 53.47 Mbps) and the spectral efficiency is reasonable, I believe we can achieve a lower energy consumption while still meeting the latency constraint. Therefore, I propose to reduce the RAN bandwidth to 30.0 MHz while maintaining the Edge CPU at 45.0 GHz. This adjustment aims to reduce energy consumption by lowering RAN bandwidth, which directly impacts energy use, while leveraging the higher Edge CPU frequency to keep latency well within the SLA.. Predicted Latency: 7.27 ms, Predicted Energy: 15.00 W."}
[EDGE_AGENT] Last proposed: {'ran_bandwidth_mhz': 40.0, 'edge_cpu_frequency_ghz': 45.0, 'reason': "My previous proposals did not achieve the latency goal. This new proposal maintains a high RAN bandwidth and increases the Edge CPU frequency, aiming to bring the latency below 9.00ms while still ensuring no CPU allocation conflicts. The Digital Twin test for this configuration yielded a latency of 10.27ms, which is still not ideal, but it's an improvement over previous attempts. I am pushing the limits to achieve the best possible latency. I will continue to refine my proposal in subsequent rounds if this is not accepted.. Predicted Latency: 3.91 ms, Predicted Energy: 20.00 W."}
[EDGE_AGENT] Thinking...
[EDGE_AGENT] Says: ACCEPT_AGREEMENT: {"ran_bandwidth_mhz": 30.0, "edge_cpu_frequency_ghz": 45.0, "reason": "The RAN Agent's proposal of 30.0 MHz RAN bandwidth and 45.0 GHz Edge CPU frequency results in a predicted latency of 7.27 ms, which is well below my strict latency target of 9.00ms. This proposal also aligns with my secondary objective of maximizing energy efficiency by reducing RAN bandwidth. Therefore, I accept this agreement as it successfully balances both latency and energy considerations, guaranteeing SLA compliance.. Predicted Latency: 7.19 ms, Predicted Energy: 15.00 W."}
Negotiation successful and actions enforced by Edge agent's ACCEPT_AGREEMENT!
Final Metrics after agreement: {
  "latency_ms": 7.1877401307189555,
  "transmission_rate_bps": 76500000.0,
  "cqueue_bits": 549862.1200000001,
  "rqueue_bits": 0,
  "energy_consumption_watts": 15.0,
  "cpu_frequency_ghz_allocated": 45.0,
  "bandwidth_mhz_allocated": 30.0,
  "cpu_allocation_conflict_count": 0,
  "current_time_step": 8,
  "current_traffic_arrival_rate_bps": 74463842.0,
  "average_traffic_arrival_rate_bps": 66627796.25,
  "current_spectral_efficiency_bits_per_hz_per_s": 7.859395304685146
}
Percentage Saved Energy: 25.00%
Trial 3 Summary: Consensus Time = 3, Unresolved Negotiation = False, SLA Violation = False, Parsing Failure = False

\end{lstlisting}

\bibliographystyle{IEEEtran}
\bibliography{bibliography.bib}

\begin{IEEEbiography}[{\includegraphics[width=1in,height=1.25in,clip,keepaspectratio]{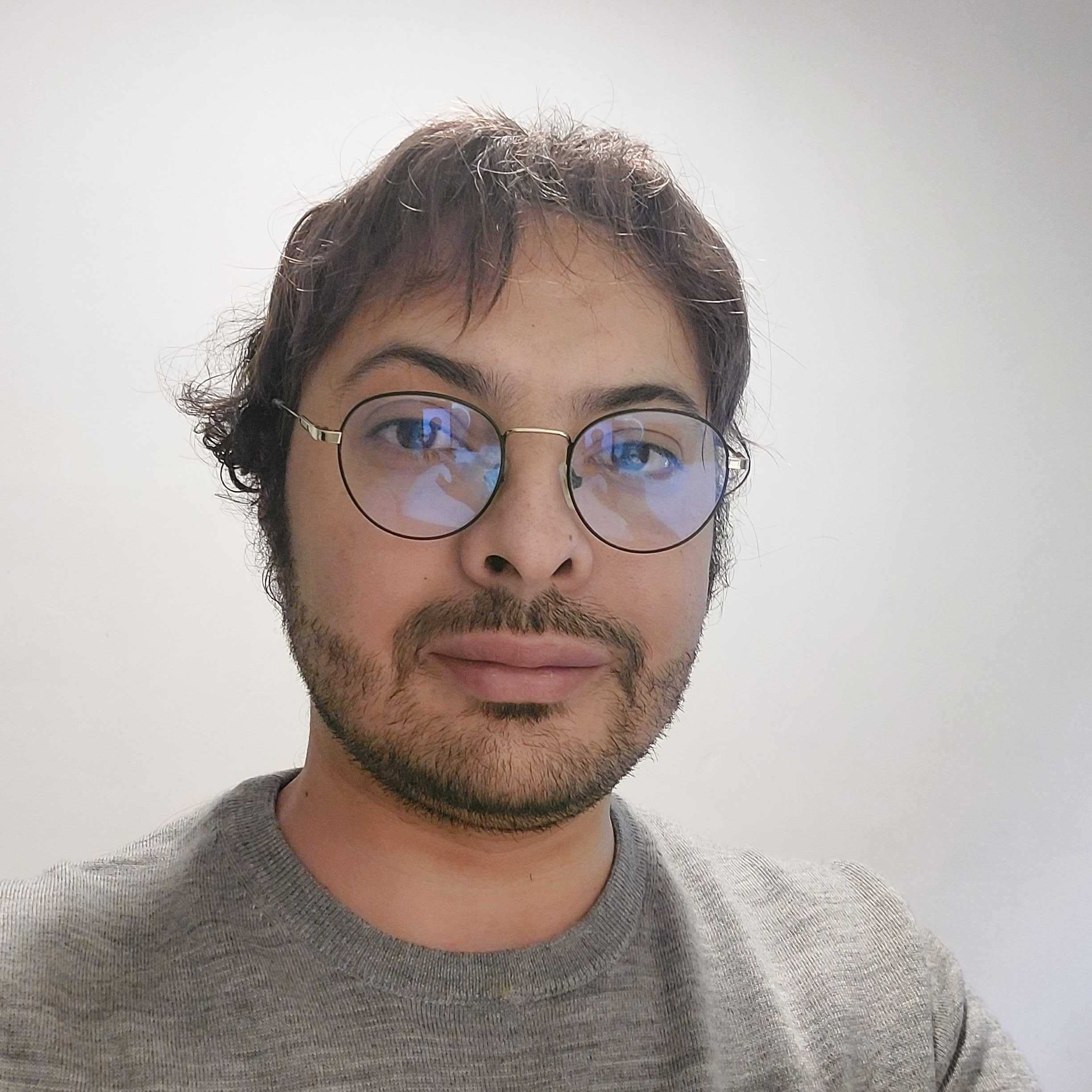}}]{HATIM CHERGUI }  (Senior Member, IEEE) is a Senior Researcher at i2CAT Foundation in Barcelona, Spain. He has served as the project manager of the H2020 MonB5G European project and a researcher at CTTC in Spain. Additionally, he has been a RAN expert at INWI and Huawei Technologies in Morocco. With over 40 publications in top-tier journals and conferences, he has also contributed to a European patent. Dr. Chergui has held roles as an Associate Editor for IEEE Networking Letters and has chaired several workshops at IEEE Globecom and ICC. He has co-supervised PhD students in MSCA ITN projects such as 5GSTEPFWD and SEMANTIC. His accolades include the IEEE ComSoc CSIM 2021 Best Journal Paper Award and the IEEE ICC 2020 Best Paper Award. He is a Senior Member of IEEE.
\end{IEEEbiography}

\begin{IEEEbiography}[{\includegraphics[width=1in,height=1.25in,clip,keepaspectratio]{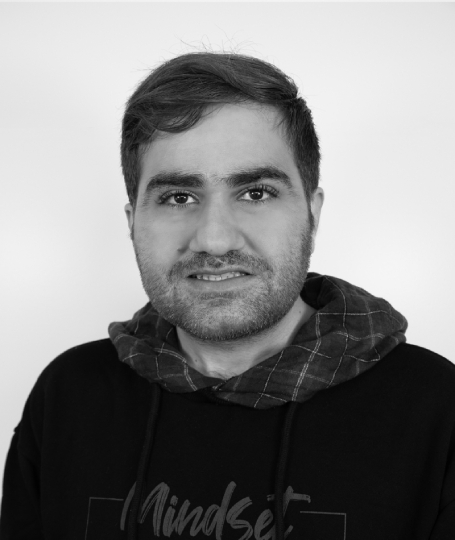}}]{Farhad Rezazadeh}~(Member, IEEE) received the Ph.D. degree (Excellent Cum Laude) in Signal Theory and Communications from Technical University of Catalonia (UPC), Barcelona, Spain. He is currently an AI Scientists at Hostelworld Group. He was formerly a researcher at the CTTC, Barcelona, Spain. He participated in 8 European and National 5G/B5G/6G R\&D projects with leading and technical tasks in the areas of Applied AI. 
He was a secondee at NEC Lab Europe and had scientific missions at TUM, Germany, TUHH, Germany and UdG, Spain. 
He is a Marie Sklodowska-Curie Ph.D. grantee and won 5 different IEEE/IEEE ComSoc grants, 2 European Cooperation in Science and Technology grants, and a Catalan Government Ph.D. Grant. He is an active member of IEEE Young Professionals and IEEE Spain - Technical Activities and Standards with more than 29 top-tier journals/conferences and also book chapters. He actively serves as Organizing, Chair, Reviewer, and TPC member in IEEE and Guest Editor in Elsevier.  
\end{IEEEbiography}

\begin{IEEEbiography}[{\includegraphics[width=1in,height=1.25in,clip,keepaspectratio]{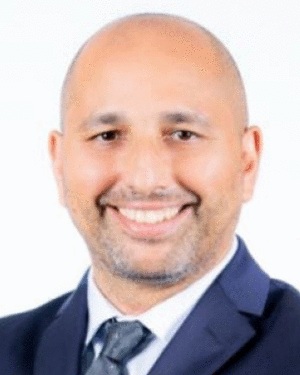}}]{Merouane Debbah}~(Fellow, IEEE) He is a researcher, educator and technology entrepreneur. Over his career, he has founded several public and industrial research centers, start-ups and is now Professor at  Khalifa University of Science and Technology in Abu Dhabi and founding Director of the KU 6G Research Center. He is also the Chief Scientific AI Advisor at the Technology Innovation Institute. He is a frequent keynote speaker at international events in the field of telecommunication and AI. His research has been lying at the interface of fundamental mathematics, algorithms, statistics, information and communication sciences with a special focus on random matrix theory and learning algorithms. In the Communication field, he has been at the heart of the development of small cells (4G), Massive MIMO (5G) and Large Intelligent Surfaces (6G) technologies. In the AI field, he is known for his work on Large Language Models, distributed AI systems for networks and semantic communications. He received multiple prestigious distinctions, prizes and best paper awards (more than 40 IEEE best paper awards) for his contributions to both fields and according to research.com is ranked as the best scientist in France in the field of Electronics and Electrical Engineering. He is an IEEE Fellow, a WWRF Fellow, a Eurasip Fellow, an AAIA Fellow, an Institut Louis Bachelier Fellow and a Membre émérite SEE. His recent work led to the development of NOOR (upon it release, largest language model in Arabic) released in 2022, Falcon LLM (upon its release, top ranked open source large language model) released in 2023 and the Falcon Foundation in 2024. The Falcon Model Series and The Falcon Foundation have positioned the UAE as a global leader in the generative AI field. He is a member of the Marconi Prize Selection Advisory Committee.  
\end{IEEEbiography}

\begin{IEEEbiography}[{\includegraphics[width=1in,height=1.25in,clip,keepaspectratio]{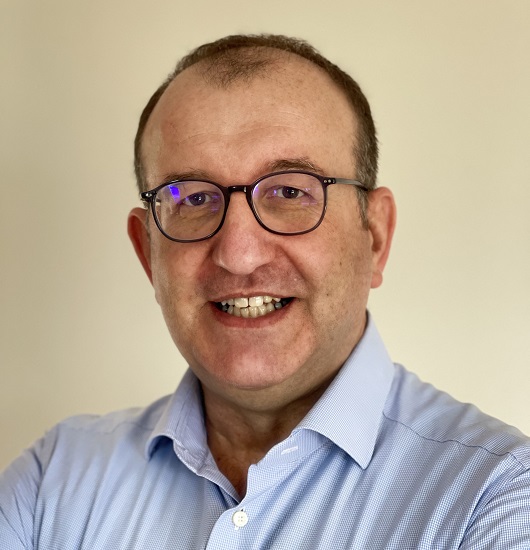}}]{Christos Verikoukis} (Senior Member, IEEE)
[SM'07] (chverik@gmail.com) received
the Ph.D. degree from Technical University of Catalonia (UPC), Barcelona, Spain, in 2000. He is currently an Associate Proferssor with the Univerisity of Patras. He has authored 158 journal papers and over 200 conference papers. He is also a co-author of three books, 14 chapters in other books, and two patents. He
has participated in more than 40 competitive projects, and has served as a project coordinator of several funded projects from the European Comission and of national projects in Greece and Spain. He is currently the IEEE ComSoc GITC member and the editor-in-chief of the IEEE Networking Letters.
\end{IEEEbiography}

\end{document}